\documentclass[aps,twocolumn,showpacs,preprintnumbers,floats]{revtex4}
\usepackage{mathrsfs}
\usepackage{longtable,lscape}
\usepackage{txfonts}
\usepackage{amssymb}
\usepackage{indentfirst}
\usepackage{graphicx,,booktabs}
\usepackage{color}
\usepackage{amssymb}
\usepackage{epsfig}
\newcommand{\vsig}{\mbox{\boldmath$\sigma$\unboldmath}}

\begin{document}
\title{Charm-strange baryon strong decays in a chiral quark model}
\author{
Lei-Hua Liu, Li-Ye Xiao and Xian-Hui Zhong \footnote {E-mail:
zhongxh@hunnu.edu.cn}} \affiliation{ Department of Physics, Hunan
Normal University, and Key Laboratory of Low-Dimensional Quantum
Structures and Quantum Control of Ministry of Education, Changsha
410081, China }


\begin{abstract}

The strong decays of charm-strange baryons up to $N=2$ shell are
studied in a chiral quark model. The theoretical predictions for the
well determined charm-strange baryons, $\Xi_c^*(2645)$,
$\Xi_c(2790)$ and $\Xi_c(2815)$, are in good agreement with the
experimental data. This model is also extended to analyze the strong
decays of the other newly observed charm-strange baryons
$\Xi_c(2930)$, $\Xi_c(2980)$, $\Xi_c(3055)$, $\Xi_c(3080)$ and
$\Xi_c(3123)$. Our predictions are given as follows. (i)
$\Xi_c(2930)$ might be the first $P$-wave excitation of $\Xi_c'$
with $J^P=1/2^-$, favors the $|\Xi_c'\ ^2P_\lambda 1/2^-\rangle$ or
$|\Xi_c'\ ^4P_\lambda 1/2^-\rangle$ state. (ii) $\Xi_c(2980)$ might
correspond to two overlapping $P$-wave states $|\Xi_c'\ ^2P_\rho
1/2^-\rangle$ and $|\Xi_c'\ ^2P_\rho 3/2^-\rangle$, respectively.
The $\Xi_c(2980)$ observed in the $\Lambda_c^+\bar{K}\pi$ final
state is most likely to be the $|\Xi_c'\ ^2P_\rho 1/2^-\rangle$
state, while the narrower resonance with a mass $m\simeq 2.97$ GeV
observed in the $\Xi_c^*(2645)\pi$ channel favors to be assigned to
the $|\Xi_c'\ ^2P_\rho 3/2^-\rangle$ state. (iii) $\Xi_c(3080)$
favors to be classified as the $|\Xi_c\ S_{\rho\rho} 1/2^+\rangle$
state, i.e., the first radial excitation ($2S$) of $\Xi_c$. (iv)
$\Xi_c(3055)$ is most likely to be the first $D$-wave excitation of
$\Xi_c$ with $J^P=3/2^+$, favors the $|\Xi_c\ ^2D_{\lambda\lambda}
3/2^+\rangle$ state. (v) $\Xi_c(3123)$ might be assigned to the
$|\Xi_c'\ ^4D_{\lambda\lambda} 3/2^+\rangle$, $|\Xi_c'\
^4D_{\lambda\lambda} 5/2^+\rangle$, or $|\Xi_c\ ^2D_{\rho\rho}
5/2^+\rangle$ state. As a by-product, we calculate the strong decays
of the bottom baryons $\Sigma_b^{\pm}$, $\Sigma_b^{*\pm}$ and
$\Xi_b^*$, which are in good agreement with the recent observations
as well.
\end{abstract}


\maketitle

\section{INTRODUCTION}

In recent years, several new charm-strange baryons, $\Xi_c(2930)$,
$\Xi_c(2980)$, $\Xi_c(3055)$, $\Xi_c(3080)$ and $\Xi_c(3123)$, have
been observed. Their experimental information has been collected in
Tab.~\ref{expdata}. $\Xi_c(2980)$ and $\Xi_c(3080)$ are relatively
well-established in experiments. Both of their isospin states were
observed by Belle Collaboration in the $\Lambda_c^+\bar{K}\pi$
channel~\cite{Chistov:2006zj}, and confirmed by BaBar with high
statistical significances~\cite{Aubert:2007dt}. Belle also observed
a resonance structure around $2.97$ GeV with a narrow width of $\sim
18$ MeV in the $\Xi_c^*(2645)\pi$ decay channel in a separate
study~\cite{Lesiak:2008wz}, which is often considered as the same
resonance, $\Xi_c(2980)$, observed in the $\Lambda_c^+\bar{K}\pi$
channel. $\Xi_c(2930)$ was found by BaBar in the $\Lambda_c^+K^-$
final state by analyzing the $B^-\rightarrow
\Lambda_c^+\bar{\Lambda}_c^-K^-$ process~\cite{Aubert:2007eb}.
However, this structure is not yet confirmed by Belle.
$\Xi_c(3055)^+$ and $\Xi_c(3123)^+$ were only observed by BaBar in
the $\Lambda_c^+K^-\pi^+$ final state with statistical significances
of 6.4$\sigma$ and $3.0\sigma$, respectively~\cite{Aubert:2007dt}.
No further evidences of them were found when BaBar searched the
inclusive $\Lambda_c^+\bar{K}$ and $\Lambda_c^+\bar{K}\pi^+\pi^-$
invariant mass spectra for new narrow states. BaBar's observations
show that $\Xi_c(3055)^+$ and $\Xi_c(3123)^+$ mostly decay though
the intermediate resonant modes $\Sigma_c(2455)^{++}K^-$ and
$\Sigma_c(2520)^{++}K^-$, respectively. A good review of the recent
experimental results on charmed baryons can be found
in~\cite{Charles:2009gr}.

Charmed baryon mass spectroscopy has been investigated in various
models~\cite{SilvestreBrac:1996bg,Migura:2006ep,Valcarce:2008dr,
Guo:2008he,Ebert:2011kk,Roberts:2007ni,Ebert:2007nw,Patel:2008mv,
Romanets:2012hm,Zhang:2008pm,Garcilazo:2007eh}. The masses of
charm-strange baryons in the $N\leq 2$ shell predicted within
several quark models have been collected in Tabs.~\ref{mass1} and
~\ref{mass2}. Comparing the experimental data with the quark model
predictions, one finds that $\Xi_c(2930)$ could be a candidate of
the $2S$ excitation of $\Xi_c$ with $J^P=1/2^+$, or the $1P$
excitation of $\Xi_c'$ with $J^P=1/2^-$, $3/2^-$ or $5/2^-$.
$\Xi_c(2980)$ might be assigned to the $2S$ excitation of $\Xi_c$ or
$\Xi_c'$ with $J^P=1/2^+$. $\Xi_c(3055)$ and $\Xi_c(3080)$ are most
likely to be the $1D$ excitations of $\Xi_c$ with $J^P=3/2^+$ or
$5/2^+$, or the $2S$ excitation of $\Xi_c'$ with $J^P=1/2^+$.
$\Xi_c(3123)$ might be classified as $1D$ excitation of $\Xi_c'$
with $J^P=3/2^+$, $5/2^+$ or $7/2^+$. Obviously, only depending on
the mass analysis it is difficult to determine the quantum numbers
of these newly observed charm-strange baryons. On the other hand,
the strong decays of these newly observed charm-strange baryons have
been studied in the framework of heavy hadron chiral perturbation
theory~\cite{Cheng:2006dk} and $^3P_0$
model~\cite{Chen:2007xf,Liu:2007ge}, respectively.
In~\cite{Cheng:2006dk}, Cheng and Chua advocated that the $J^P$
numbers of $\Xi_c(2980)$ and $\Xi_c(3080)$ could be $1/2^+$ and
$5/2^+$, respectively. They claimed that under this $J^P$
assignment, it is easy to understand why $\Xi_c(2980)$ is broader
than $\Xi_c(3080)$. In ~\cite{Chen:2007xf,Liu:2007ge}, Chen \emph{et
al.} have analyzed the strong decays of the $N=2$ shell excited
charm-strange baryons in the $^3P_0$ model, they could only exclude
some assignments according to the present experimental information.
As a whole, although the new charm-strange baryons have been studied
in several aspects, such as mass spectroscopy and strong decays,
their quantum numbers are not clear so far. Thus, more
investigations of these new heavy baryons are needed.

To further understand the nature of these newly observed
charm-strange baryons, in this work, we make a systematic study of
their strong decays in a chiral quark model, which has been
developed and successfully used to deal with the strong decays of
charmed baryons and heavy-light
mesons~\cite{Zhong:2007gp,Zhong:2010vq,Zhong:2009sk,Zhong:2008kd}.
It should be pointed out that very recently, some important
progresses in the observation of the bottom baryons have been
achieved in experiments as well: CDF Collaboration first measured
the natural widths of the bottom baryons $\Sigma_b^{\pm}$ and
$\Sigma_b^{*\pm}$, and improved the measurement
masses~\cite{CDF:2011ac}, and CMS Collaboration observed a new
neutral excited bottom baryon with a mass $m=5945.0\pm 0.7\pm 0.3\pm
2.7$ MeV, which is most likely to be the $\Xi_b^{*0}$~\cite{CMS}. As
a by-product, in this work we also calculate the strong decays of
these bottom baryons according to the new measurements.

This work is organized as follows. In the subsequent section, the
charm-strange baryon in the quark model is outlined. Then a brief
review of the chiral quark model approach is given in
Sec.~\ref{qmc}. The numerical results are presented and discussed in
Sec.~\ref{rds}. Finally, a summary is given in Sec.~\ref{sumary}.

\begin{widetext}
\begin{center}
\begin{table}[ht]
\caption{Summary of the experimental results of the newly observed
charm-strange baryons.} \label{expdata}
\begin{tabular}{|c|c|c|c|c|c|c }\hline\hline
Resonance & Mass (MeV)& Width (MeV)&  Observed decay channel & Collaboration & Status~\cite{Nakamura:2010zzi}  \\
\hline
$\Xi_c(2930)^0$ & $2931\pm3\pm 5 $ & $36\pm 7\pm 11$ &$\Lambda_c^+K^-$  & BaBar~\cite{Aubert:2007eb} & *     \\
\hline
$\Xi_c(2980)^+$ & $2978.5\pm 2.1\pm 2.0$ & $43.5\pm 7.5\pm7.0$ &$\Lambda_c^+K^-\pi^+$ & Belle~\cite{Chistov:2006zj} & ***    \\
                & $2969.3\pm 2.2\pm 1.7$ & $27\pm 8\pm2$ &$\Lambda_c^+K^-\pi^+$,\ $\Sigma_c(2455)^{++}K^-$   & BaBar~\cite{Aubert:2007dt}&   \\
                & $2967.7\pm 2.3^ {+1.1}_{-1.2}$ & $18\pm6\pm3$ &$\Xi_c(2645)^0\pi^+$ & Belle~\cite{Lesiak:2008wz} &    \\
\hline
$\Xi_c(2980)^0$ & $2977.1\pm 8.8\pm 3.5$ & $43.5$ fixed &$\Lambda_c^+\bar{K}^0\pi^-$  & Belle~\cite{Chistov:2006zj} & ***   \\
                & $2972.9\pm 4.4\pm 1.6$ & $31\pm 7\pm8$ &$\Lambda_c^+\bar{K}^0\pi^-$  & BaBar~\cite{Aubert:2007dt} &   \\
                & $2965.7\pm 2.4^ {+1.1}_{-1.2}$ & $15\pm6\pm3$   &$\Xi_c(2645)^+\pi^-$ &Belle ~\cite{Lesiak:2008wz}&    \\
\hline
$\Xi_c(3055)^+$ & $3054.2\pm 1.2\pm 0.5$ & $17\pm 6\pm 11$ &$\Sigma_c(2455)^{++}K^-$ & BaBar~\cite{Aubert:2007dt}& **\\
\hline
$\Xi_c(3080)^+$ & $3076.7\pm 0.9\pm 0.5$ & $6.2\pm 1.2\pm0.8$ &$\Lambda_c^+K^-\pi^+$ & Belle~\cite{Chistov:2006zj}&***     \\
                & $3077.0\pm 0.4\pm 0.2$ & $5.5\pm 1.3\pm0.6$ &$\Lambda_c^+K^-\pi^+$,\ $\Sigma_c(2455,2520)^{++}K^-$ &BaBar~\cite{Aubert:2007dt} &   \\
\hline
$\Xi_c(3080)^0$ & $3082.8\pm 1.8\pm 1.5$ & $5.2\pm 3.1\pm1.8$ &$\Lambda_c^+\bar{K}^0\pi^-$  &Belle~\cite{Chistov:2006zj}&***    \\
                & $3079.3\pm 1.1\pm 0.2$ & $5.9\pm 2.3\pm1.5$ &$\Lambda_c^+\bar{K}^0\pi^-$, $\Sigma_c(2455,2520)^{0}K^0_S$  & BaBar~\cite{Aubert:2007dt}&   \\
\hline
$\Xi_c(3123)^+$ & $3122.9\pm 1.3\pm 0.3$ & $4.4\pm 3.4\pm 1.7$ &$\Sigma_c(2520)^{++}K^-$ & BaBar~\cite{Aubert:2007dt}& *\\
\hline
\end{tabular}
\end{table}
\end{center}
\end{widetext}

\begin{table}[ht]
\caption{Masses (MeV) of the $\Xi_c$-type charm-strange baryons in
the various quark models. } \label{mass1}
\begin{tabular}{|c |c|c|c |c|c|c|c|c}\hline\hline
N &  $J^P$ & State & \cite{Valcarce:2008dr} & \cite{Ebert:2011kk}&\cite{SilvestreBrac:1996bg}&\cite{Roberts:2007ni}& \cite{Ebert:2007nw}  \\
\hline
0 &  $\frac{1}{2}^+$ &1S & 2471 & 2476 & 2467  &2466 & 2481  \\
1 &  $\frac{1}{2}^-$ & 1P& 2799 & 2792 & 2792  &2773 & 2801 \\
1 &  $\frac{3}{2}^-$ &1P & ...     & 2819 &2792   &2783 & 2820 \\
2 &  $\frac{1}{2}^+$ &2S & 3137 &  2959& 2992  &...   & 2923 \\
2 &  $\frac{3}{2}^+$ &1D & 3071 &  3059& 3057  &3012 & 3030 \\
2 &  $\frac{5}{2}^+$ &1D & 3049 &  3076& 3057  &3004 & 3042 \\
\hline
\end{tabular}
\end{table}

\begin{table}[ht]
\caption{Masses (MeV) of the $\Xi_c'$-type charm-strange baryons in
the various quark models.} \label{mass2}
\begin{tabular}{|c|c|c|c|c|c|c|c|c}\hline\hline
N &  $J^P$  & State &\cite{Valcarce:2008dr} & \cite{Ebert:2011kk}&\cite{SilvestreBrac:1996bg}&\cite{Roberts:2007ni}& \cite{Ebert:2007nw}  \\
\hline
0 &  $\frac{1}{2}^+$ &1S & 2574 & 2579 & 2567 & 2594  & 2578 \\
0 &  $\frac{3}{2}^+$ &1S & 2642 & 2649 & 2647 &2649   & 2654\\
1 &  $\frac{1}{2}^-$ &1P & 2902 & 2936 & 2897 &2855   & 2934 \\
1 &  $\frac{3}{2}^-$ &1P & ...     & 2935 & 2910   &2866 & 2931 \\
1 &  $\frac{5}{2}^-$ &1P & ...   & 2929   &3050   &2895 & 2921 \\
2 &  $\frac{1}{2}^+$ &2S & 3212 & 2983   & 3087  & ...   & 2984\\
2 &  $\frac{1}{2}^+$ &1D & ... & 3163   & ... & ...  & 3132 \\
2 &  $\frac{3}{2}^+$ &1D & ...   & 3160& 3127   &... & 3127\\
2 &  $\frac{5}{2}^+$ &1D & 3132 &  3166& 3167  &3080  & 3123\\
2 &  $\frac{7}{2}^+$ &1D & ...     &  3147    & ...      &3094 & 3136\\
\hline
\end{tabular}
\end{table}

\section{charm-strange baryon in the quark model}\label{spctr}

The charmed baryon contains a heavy charm quark, which violates
SU(4) symmetry. However, the SU(3) symmetry between the other two
light quarks ($u$, $d$, or $s$) is approximately kept. According to
the symmetry, the charmed baryons can be classified two different
SU(3) flavor representations: the symmetric $\mathbf{6}$ and
antisymmetric antitriplet $\bar{\mathbf{3}}$. For the charm-strange
baryon, the antisymmetric flavor wave function ($\Xi_c$-type) can be
written as
\begin{eqnarray}
\phi_{\Xi_c}=\cases{\frac{1}{\sqrt{2}}(us-su)c & for $\Xi^{+}_c$\cr
\frac{1}{\sqrt{2}}(ds-sd)c  & for $\Xi^{0}_c$},
\end{eqnarray}
while the symmetric flavor wave function ($\Xi'_c$-type) is given by
\begin{eqnarray}
\phi_{\Xi'_c}=\cases{\frac{1}{\sqrt{2}}(u s+su) c & for
$\Xi'^{+}_c$\cr \frac{1}{\sqrt{2}}(ds+sd)c  & for $\Xi'^{0}_c$}.
\end{eqnarray}

In the quark model, the typical SU(2) spin wave functions for the
charm-strange baryons can be adopted
\cite{Koniuk:1979vy,Copley:1979wj}, which are
\begin{eqnarray}
\chi^s_{3/2}&=&\uparrow\uparrow\uparrow, \ \  \chi^s_{-3/2}=\downarrow\downarrow\downarrow, \nonumber\\
\chi^s_{1/2}&=&\frac{1}{\sqrt{3}}(\uparrow\uparrow\downarrow+\uparrow\downarrow\uparrow+\downarrow\uparrow\uparrow),\nonumber\\
\chi^s_{-1/2}&=&\frac{1}{\sqrt{3}}(\uparrow\downarrow\downarrow+\downarrow\downarrow\uparrow+\downarrow\uparrow\downarrow),
\end{eqnarray}
for the spin-3/2 states with a symmetric spin wave function,
\begin{eqnarray}
\chi^\rho_{1/2}&=&\frac{1}{\sqrt{2}}(\uparrow\downarrow\uparrow-\downarrow\uparrow\uparrow),\nonumber\\
\chi^\rho_{-1/2}&=&\frac{1}{\sqrt{2}}(\uparrow\downarrow\downarrow-\downarrow\uparrow\downarrow),
\end{eqnarray}
for the spin-1/2 states with a mixed antisymmetric spin wave
function, and
\begin{eqnarray}
\chi^\lambda_{1/2}&=&-\frac{1}{\sqrt{6}}(\uparrow\downarrow\uparrow+\downarrow\uparrow\uparrow-2\uparrow\uparrow\downarrow),\nonumber\\
\chi^\lambda_{-1/2}&=&\frac{1}{\sqrt{6}}(\uparrow\downarrow\downarrow+\downarrow\uparrow\downarrow-2\downarrow\downarrow\uparrow),
\end{eqnarray}
for the spin-1/2 states with a mixed symmetric spin wave function.

The spatial wave function of a charm-strange baryon is adopted the
harmonic oscillator form in the constituent quark model. The details
of the spatial wave functions can be found in our previous
work~\cite{Zhong:2007gp}.

The spin-flavor and spatial wave functions of baryons must be
symmetric since the color wave function is antisymmetric. The flavor
wave functions of the $\Xi_c$-type charm-strange baryons,
$\phi_{\Xi_c}$, are antisymmetric under the interchange of the $u$
($d$) and $s$ quarks, thus, their spin-space wave functions must be
symmetric. In contrast, the spin-spatial wave functions of
$\Xi'_c$-type charm-strange baryons are required to be antisymmetric
due to their symmetric flavor wave functions under the interchange
of the two light quarks. The notations, wave functions, and quantum
numbers of the $\Xi_c$-type and $\Xi'_c$-type charm-strange baryons
up to $N=2$ shell classified in the quark model are listed in Tabs.
\ref{wf1} and \ref{wf2}, respectively.

\begin{widetext}
\begin{center}
\begin{table}[ht]
\caption{The $\Xi_c$-type charm-strange baryons classified in the
quark model and their possible two body strong decay channels. The
notation of the $\Xi_c$-type charmed baryon is denoted by $|\Xi_c \
^{2S+1} L_{\sigma}J^P\rangle$ as used in Ref.~\cite{Copley:1979wj}.
The Clebsch-Gordan series for the spin and angular-momentum addition
$|\Xi_c \ ^{2S+1} L_{\sigma}J^P\rangle= \sum_{L_z+S_z=J_z} \langle
LL_z,SS_z|JJ_z \rangle  ^{N}\Psi^{\sigma }_{LL_z}
\chi_{S_z}\phi_{\Xi_c}$ has been omitted. } \label{wf1}
\begin{tabular}{|c |  c  c  c  c  c  c |c|c|c|c|c|c }\hline\hline
Notation & N  &$l_\lambda$ & $l_\rho$ & \ L & $S$  & \ \ $J^P$  \ \ &\ \ Wave function & Strong decay channel \\
\hline
$|\Xi_c \ ^2 S\frac{1}{2}^+\rangle$                  &0& 0 & 0& 0 & $\frac{1}{2}$     & $\frac{1}{2}^+$  & $^{0}\Psi^S_{00}\chi^\rho_{S_z}\phi_{\Xi_c}$ & ...      \\
\hline
$|\Xi_c \ ^2 P_\lambda\frac{1}{2}^-\rangle$          &1& 1 &0 & 1 & $\frac{1}{2}$  &$\frac{1}{2}^-$ &  $   ^{1}\Psi^{\lambda}_{1L_z} \chi^{\rho}_{S_z}\phi_{\Xi_c}$& $\Xi_c'\pi$, $\Xi_c^*\pi$                      \\
$|\Xi_c \ ^2 P_\lambda\frac{3}{2}^-\rangle$          &1& 1 & 0& 1 & $\frac{1}{2}$  &$\frac{3}{2}^-$ &      &   \\
\hline
$|\Xi_c \ ^2 P_\rho\frac{1}{2}^-\rangle$             &1& 0 & 1& 1 & $\frac{1}{2}$  &$\frac{1}{2}^-$ &  $  ^{1}\Psi^{\rho}_{1L_z} \chi^{\lambda}_{S_z}\phi_{\Xi_c}$ &$\Xi_c\pi$, $\Xi_c'\pi$, $\Xi_c^*\pi$       \\
$|\Xi_c \ ^2 P_\rho\frac{3}{2}^-\rangle$             &1& 0 & 1& 1 & $\frac{1}{2}$  &$\frac{3}{2}^-$ &   &     \\
\hline
$|\Xi_c \ ^4 P_\rho\frac{1}{2}^-\rangle$             &1& 0 & 1& 1 & $\frac{3}{2}$  &$\frac{1}{2}^-$ &  &      \\
$|\Xi_c \ ^4 P_\rho\frac{3}{2}^-\rangle$             &1& 0 & 1& 1 & $\frac{3}{2}$  &$\frac{3}{2}^-$ &  $   ^{1}\Psi^{\rho}_{1L_z} \chi^{S}_{S_z}\phi_{\Xi_c}$ &$\Xi_c\pi$, $\Xi_c'\pi$, $\Xi_c^*\pi$     \\
$|\Xi_c \ ^4 P_\rho\frac{5}{2}^-\rangle$             &1& 0 &1 & 1 & $\frac{3}{2}$  &$\frac{5}{2}^-$ & &\\
\hline
$|\Xi_c \ ^2 S_A \frac{1}{2}^+\rangle$               &2& 1 & 1& 0 & $\frac{1}{2}$  &$\frac{1}{2}^+$   & $^{2}\Psi^{A}_{00}\chi^\lambda_{S_z}\phi_{\Xi_c}$ & $\Xi_c\pi$,$\Xi_c\eta$,$\Lambda_c K$, $\Xi_c'\pi$,$\Sigma_cK$, $\Xi_c^*\pi$,$\Sigma_c^*K$,$\Xi_c(2790,2815)\pi$  \\
\hline
$|\Xi_c \ ^4 S_A \frac{3}{2}^+\rangle$               &2& 1 & 1& 0 & $\frac{3}{2}$  &$\frac{3}{2}^+$   & $^{2}\Psi^{A}_{00}\chi^s_{S_z}\phi_{\Xi_c}$& $\Xi_c\pi$, $\Xi_c\eta$,$\Lambda_c K$, $\Xi_c'\pi$,$\Sigma_cK$, $\Xi_c^*\pi$,$\Sigma_c^*K$,$\Xi_c(2790,2815)\pi$      \\
\hline
$|\Xi_c \ ^2 P_A \frac{1}{2}^+\rangle$               &2& 1 & 1& 1 & $\frac{1}{2}$    &$\frac{1}{2}^+$   &  $^{2}\Psi^{A}_{1L_z} \chi^{\lambda}_{S_z}\phi_{\Xi_c}$  & $\Xi_c\pi$,$\Xi_c\eta$, $\Lambda_c K$, $\Xi_c'\pi$,$\Sigma_cK$, $\Xi_c^*\pi$,$\Sigma_c^*K$,$\Xi_c(2790,2815)\pi$      \\
$|\Xi_c \ ^2 P_A \frac{3}{2}^+\rangle$               &2& 1 & 1& 1 & $\frac{1}{2}$    &$\frac{3}{2}^+$   &  &        \\
\hline
$|\Xi_c \ ^4 P_A\frac{1}{2}^+\rangle$                &2& 1 &1 & 1 & $\frac{3}{2}$  &$\frac{1}{2}^+$ & & $\Xi_c\pi$,$\Xi_c\eta$, $\Lambda_c K$, $\Xi_c'\pi$,$\Sigma_cK$, $\Xi_c^*\pi$,$\Sigma_c^*K$,$\Xi_c(2790,2815)\pi$       \\
$|\Xi_c \ ^4 P_A\frac{3}{2}^+\rangle$                &2& 1 &1 & 1 & $\frac{3}{2}$   &$\frac{3}{2}^+$ &$  \ \ ^{2}\Psi^{A}_{1L_z} \chi^{S}_{S_z}\phi_{\Xi_c}$ & $\Xi_c\pi$,$\Xi_c\eta$,$\Lambda_c K$, $\Xi_c'\pi$,$\Sigma_cK$, $\Xi_c^*\pi$,$\Sigma_c^*K$,$\Xi_c(2790,2815)\pi$       \\
$|\Xi_c \ ^4 P_A\frac{5}{2}^+\rangle$                &2& 1 &1 & 1 & $\frac{3}{2}$    &$\frac{5}{2}^+$ &  & $\Xi_c\pi$, $\Xi_c\eta$,$\Lambda_c K$, $\Xi_c^*\pi$,$\Sigma_c^*K$,$\Xi_c(2790,2815)\pi$         \\
\hline
$|\Xi_c \ ^2 D_A\frac{3}{2}^+\rangle$                &2& 1 & 1& 2 & $\frac{1}{2}$  &$\frac{3}{2}^+$  &   &$\Xi_c\pi$,$\Xi_c\eta$, $\Lambda_c K$, $\Xi_c'\pi$,$\Sigma_cK$,$\Xi_c^*\pi$,$\Sigma_c^*K$,$\Xi_c(2790,2815)\pi$   \\
$|\Xi_c \ ^2 D_A\frac{5}{2}^+\rangle$                &2& 1 & 1& 2 & $\frac{1}{2}$    &$\frac{5}{2}^+$    & $  \ \ ^{2}\Psi^{A}_{2L_z} \chi^{\lambda}_{S_z}\phi_{\Xi_c}$   &     \\
\hline
$|\Xi_c \ ^4 D_A\frac{1}{2}^+\rangle$                &2& 1 & 1& 2 & $\frac{3}{2}$     &$\frac{1}{2}^+$  &    &$\Xi_c\pi$,$\Xi_c\eta$, $\Lambda_c K$, $\Xi_c'\pi$,$\Sigma_cK$,$\Xi_c^*\pi$,$\Sigma_c^*K$,$\Xi_c(2790,2815)\pi$     \\
$|\Xi_c \ ^4 D_A\frac{3}{2}^+\rangle$                &2& 1 & 1& 2 & $\frac{3}{2}$   &$\frac{3}{2}^+$   &    &   \\
$|\Xi_c \ ^4 D_A\frac{5}{2}^+\rangle$                &2& 1 & 1& 2 & $\frac{3}{2}$    &$\frac{5}{2}^+$   &$  \ \ ^{2}\Psi^{A}_{2L_z} \chi^{S}_{S_z}\phi_{\Xi_c}$   &     \\
$|\Xi_c \ ^4 D_A\frac{7}{2}^+\rangle$                &2& 1 & 1& 2 & $\frac{3}{2}$     &$\frac{7}{2}^+$   &  &        \\
\hline
$|\Xi_c \ ^2 D_{\rho\rho}\frac{3}{2}^+\rangle$       &2& 0 & 2& 2 & $\frac{1}{2}$   &$\frac{3}{2}^+$    & &$\Xi_c'\pi$,$\Sigma_cK$,$\Xi_c^*\pi$,$\Sigma_c^*K$ \\
$|\Xi_c \ ^2 D_{\rho\rho}\frac{5}{2}^+\rangle$       &2& 0 & 2& 2 & $\frac{1}{2}$     &$\frac{5}{2}^+$   & $  \ ^{2}\Psi^{\rho\rho}_{2L_z} \chi^{\rho}_{S_z}\phi_{\Xi_c}$& \\
 \hline
$|\Xi_c \ ^2 D_{\lambda\lambda}\frac{3}{2}^+\rangle$ &2&  2& 0& 2 & $\frac{1}{2}$   &$\frac{3}{2}^+$ &     &$\Xi_c'\pi$,$\Sigma_cK$,$\Xi_c^*\pi$,$\Sigma_c^*K$,$D\Lambda$      \\
$|\Xi_c \ ^2 D_{\lambda\lambda}\frac{5}{2}^+\rangle$ &2&  2& 0& 2 & $\frac{1}{2}$    &$\frac{5}{2}^+$ & $  \ \    ^{2}\Psi^{\lambda\lambda}_{2L_z} \chi^{\rho}_{S_z}\phi_{\Xi_c}$&\\
\hline
$|\Xi_c \ ^2 S_{\rho\rho}\frac{1}{2}^+\rangle$       &2&  0& 0& 0 & $\frac{1}{2}$    &$\frac{1}{2}^+$ &$^{2}\Psi^{\rho\rho}_{00} \chi^{\rho}_{S_z}\phi_{\Xi_c}$     &$\Xi_c'\pi$,$\Sigma_cK$,$\Xi_c^*\pi$,$\Sigma_c^*K$  \\
$|\Xi_c \ ^2 S_{\lambda\lambda}\frac{1}{2}^+\rangle$ &2&  0& 0& 0 & $\frac{1}{2}$   &$\frac{1}{2}^+$ &$^{2}\Psi^{\lambda\lambda}_{00} \chi^{\rho}_{S_z}\phi_{\Xi_c}$ &$\Xi_c'\pi$,$\Sigma_cK$,$\Xi_c^*\pi$,$\Sigma_c^*K$,$D\Lambda$\\
\hline
\end{tabular}
\end{table}
\end{center}
\end{widetext}

\begin{widetext}
\begin{center}
\begin{table}[ht]
\caption{The $\Xi_c'$-type charm-strange baryons classified in the
quark model and their possible two body strong decay channels. The
notation of the $\Xi_c'$-type charmed baryon is denoted by $|\Xi_c'
\ ^{2S+1} L_{\sigma}J^P\rangle$ as used in
Ref.~\cite{Copley:1979wj}. The Clebsch-Gordan series for the spin
and angular-momentum addition $|\Xi_c' \ ^{2S+1}
L_{\sigma}J^P\rangle= \sum_{L_z+S_z=J_z} \langle LL_z,SS_z|JJ_z
\rangle  ^{N}\Psi^{\sigma }_{LL_z} \chi_{S_z}\phi_{\Xi_c'}$ has been
omitted.} \label{wf2}
\begin{tabular}{|c |  c  c  c  c  c  c |c|c|c|c|c|c }\hline\hline
Notation & $N$  & $l_\lambda$& $l_\rho$  & $L$ & $S$ & \ \ $J^P$  \ \ &\ \ Wave function & Strong decay channel  \\
\hline
$|\Xi_c^{'} \ ^2 S\frac{1}{2}^+\rangle$                  & 0 &0&0    & 0 & $\frac{1}{2}$ & $\frac{1}{2}^+$  & $^{0}\Psi^S_{00}\chi^\lambda_{S_z}\phi_{\Xi_c^{'}}$               & ...    \\
\hline
$|\Xi_c^{'} \ ^4 S\frac{3}{2}^+\rangle$                  & 0 &0&0    & 0 & $\frac{3}{2}$ & $\frac{3}{2}^+$  & $^{0}\Psi^S_{00}\chi^s_{S_z}\phi_{\Xi_c^{'}}$                     & $\Xi_c\pi$    \\
\hline
$|\Xi_c^{'} \ ^2 P_\lambda\frac{1}{2}^-\rangle$          & 1 &1&0    & 1 & $\frac{1}{2}$ & $\frac{1}{2}^-$   &  $^{1}\Psi^{\lambda}_{1L_z} \chi^{\lambda}_{S_z}\phi_{\Xi_c^{'}}$ &$\Xi_c\pi$, $\Lambda_cK$,  $\Xi_c'\pi$, $\Sigma_cK$,$\Xi_c^*\pi$, $\Sigma_c^*K$,$\Xi_c(2790,2815)\pi$                \\
$|\Xi_c^{'} \ ^2 P_\lambda\frac{3}{2}^-\rangle$          & 1 &1&0    & 1 & $\frac{1}{2}$ & $\frac{3}{2}^-$   &   &      \\
\hline
$|\Xi_c^{'} \ ^4 P_\lambda\frac{1}{2}^-\rangle$          & 1 &1&0    & 1 & $\frac{3}{2}$ &$\frac{1}{2}^-$   &   &$\Xi_c\pi$, $\Lambda_cK$,  $\Xi_c'\pi$, $\Sigma_cK$,$\Xi_c^*\pi$, $\Sigma_c^*K$,$\Xi_c(2790,2815)\pi$                      \\
$|\Xi_c^{'} \ ^4 P_\lambda\frac{3}{2}^-\rangle$          & 1 &1&0    & 1 & $\frac{3}{2}$ &$\frac{3}{2}^-$ &  $   ^{1}\Psi^{\lambda}_{1L_z} \chi^{s}_{S_z}\phi_{\Xi_c^{'}}$ &\\
$|\Xi_c^{'} \ ^4 P_\lambda\frac{5}{2}^-\rangle$          & 1 &1&0    & 1 & $\frac{3}{2}$ &$\frac{5}{2}^-$ &                 &     \\
\hline
$|\Xi_c^{'} \ ^2 P_\rho\frac{1}{2}^-\rangle$             & 1 &0&1    & 1 & $\frac{1}{2}$ &$\frac{1}{2}^-$  &  $^{1}\Psi^{\rho}_{1L_z} \chi^{\rho}_{S_z}\phi_{\Xi_c^{'}}$ & $\Xi_c'\pi$, $\Sigma_cK$,$\Xi_c^*\pi$, $\Sigma_c^*K$        \\
$|\Xi_c^{'} \ ^2 P_\rho\frac{3}{2}^-\rangle$             & 1 &0&1    & 1 & $\frac{1}{2}$ &$\frac{3}{2}^-$ &    &    \\
\hline
$|\Xi_c^{'} \ ^2 S_A \frac{1}{2}^+\rangle$               & 2 &1&1    & 0 & $\frac{1}{2}$ &$\frac{1}{2}^+$   & $^{2}\Psi^{A}_{00}\chi^\rho_{S_z}\phi_{\Xi_c^{'}}$ &$\Xi_c'\pi$, $\Sigma_cK$,$\Xi_c^*\pi$, $\Sigma_c^*K$    \\
\hline
$|\Xi_c^{'} \ ^2 P_A \frac{1}{2}^+\rangle$               & 2 &1&1    & 1 & $\frac{1}{2}$ &$\frac{1}{2}^+$   &  $ \ \ ^{2}\Psi^{A}_{1L_z} \chi^{\rho}_{S_z}\phi_{\Xi_c^{'}}$ &$\Xi_c'\pi$, $\Sigma_cK$,$\Xi_c^*\pi$, $\Sigma_c^*K$        \\
$|\Xi_c^{'}\ ^2 P_A \frac{3}{2}^+\rangle$                & 2 &1&1    & 1 & $\frac{1}{2}$ &$\frac{3}{2}^+$   &     &     \\
\hline
$|\Xi_c^{'} \ ^2 D_A\frac{3}{2}^+\rangle$                & 2 &1&1    & 2 & $\frac{1}{2}$ &$\frac{3}{2}^+$  &   &$\Xi_c'\pi$, $\Sigma_cK$,$\Xi_c^*\pi$, $\Sigma_c^*K$    \\
$|\Xi_c^{'} \ ^2 D_A\frac{5}{2}^+\rangle$                & 2 &1&1    & 2 & $\frac{1}{2}$ &$\frac{5}{2}^+$    & $  \ \ ^{2}\Psi^{A}_{2L_z} \chi^{\rho}_{S_z}\phi_{\Xi_c^{'}}$  &      \\
\hline
$|\Xi_c^{'} \ ^2 D_{\rho\rho}\frac{3}{2}^+\rangle$       & 2 &0&2    & 2 & $\frac{1}{2}$ &$\frac{3}{2}^+$    & &$\Xi_c\pi$, $\Xi_c\eta$, $\Lambda_cK$,  $\Xi_c'\pi$, $\Sigma_cK$,$\Xi_c^*\pi$, $\Sigma_c^*K$,$\Xi_c(2790,2815)\pi$    \\
$|\Xi_c^{'} \ ^2 D_{\rho\rho}\frac{5}{2}^+\rangle$       & 2 &0&2    & 2 & $\frac{1}{2}$ &$\frac{5}{2}^+$   & $  \ ^{2}\Psi^{\rho\rho}_{2L_z} \chi^{\lambda}_{S_z}\phi_{\Xi_c^{'}}$ & \\
\hline
$|\Xi_c^{'} \ ^4 D_{\rho\rho}\frac{1}{2}^+\rangle$       & 2 &0&2    & 2 & $\frac{3}{2}$ &$\frac{1}{2}^+$    & &$\Xi_c\pi$,$\Xi_c\eta$, $\Lambda_cK$,  $\Xi_c'\pi$, $\Sigma_cK$,$\Xi_c^*\pi$, $\Sigma_c^*K$,$\Xi_c(2790,2815)\pi$   \\
$|\Xi_c^{'} \ ^4 D_{\rho\rho}\frac{3}{2}^+\rangle$       & 2 &0&2    & 2 & $\frac{3}{2}$ &$\frac{3}{2}^+$   & $  \ ^{2}\Psi^{\rho\rho}_{2L_z} \chi^{s}_{S_z}\phi_{\Xi_c^{'}}$& \\
$|\Xi_c^{'} \ ^4 D_{\rho\rho}\frac{5}{2}^+\rangle$       & 2 &0&2    & 2 & $\frac{3}{2}$ &$\frac{5}{2}^+$    &  &     \\
$|\Xi_c^{'} \ ^4 D_{\rho\rho}\frac{7}{2}^+\rangle$       & 2 &0&2    & 2 & $\frac{3}{2}$ &$\frac{7}{2}^+$   &  &   \\
\hline
$|\Xi_c^{'} \ ^2 D_{\lambda\lambda}\frac{3}{2}^+\rangle$ & 2 &2&0    & 2 & $\frac{1}{2}$ &$\frac{3}{2}^+$ & &$\Xi_c\pi$,$\Xi_c\eta$, $\Lambda_cK$,  $\Xi_c'\pi$, $\Sigma_cK$,$\Xi_c^*\pi$, $\Sigma_c^*K$,$\Xi_c(2790,2815)\pi$,$D\Lambda$           \\
$|\Xi_c^{'} \ ^2 D_{\lambda\lambda}\frac{5}{2}^+\rangle$ & 2 &2&0    & 2 & $\frac{1}{2}$ &$\frac{5}{2}^+$ & $ \ \    ^{2}\Psi^{\lambda\lambda}_{2L_z} \chi^{\lambda}_{S_z}\phi_{\Xi_c^{'}}$ & \\
\hline
$|\Xi_c^{'} \ ^4 D_{\lambda\lambda}\frac{1}{2}^+\rangle$ & 2 &2&0    & 2 & $\frac{3}{2}$ &$\frac{1}{2}^+$    &&$\Xi_c\pi$,$\Xi_c\eta$, $\Lambda_cK$,  $\Xi_c'\pi$, $\Sigma_cK$,$\Xi_c^*\pi$, $\Sigma_c^*K$,$\Xi_c(2790,2815)\pi$,$D\Lambda$    \\
$|\Xi_c^{'} \ ^4 D_{\lambda\lambda}\frac{3}{2}^+\rangle$ & 2 &2&0    & 2 & $\frac{3}{2}$ &$\frac{3}{2}^+$   & $  \ ^{2}\Psi^{\lambda\lambda}_{2L_z} \chi^{s}_{S_z}\phi_{\Xi_c^{'}}$& \\
$|\Xi_c^{'} \ ^4 D_{\lambda\lambda}\frac{5}{2}^+\rangle$ & 2 &2&0    & 2 & $\frac{3}{2}$ &$\frac{5}{2}^+$    &   &    \\
$|\Xi_c^{'} \ ^4 D_{\lambda\lambda}\frac{7}{2}^+\rangle$ & 2 &2&0    & 2 & $\frac{3}{2}$ &$\frac{7}{2}^+$   &   &  \\
\hline
$|\Xi_c^{'} \ ^2 S_{\rho\rho}\frac{1}{2}^+\rangle$       & 2 &0&0    & 0 & $\frac{1}{2}$ &$\frac{1}{2}^+$ & $^{2}\Psi^{\rho\rho}_{00} \chi^{\lambda}_{S_z}\phi_{\Xi_c^{'}}$ &$\Xi_c\pi$,$\Xi_c\eta$, $\Lambda_cK$,  $\Xi_c'\pi$, $\Sigma_cK$,$\Xi_c^*\pi$, $\Sigma_c^*K$,$\Xi_c(2790,2815)\pi$      \\
\hline
$|\Xi_c^{'} \ ^4 S_{\rho\rho}\frac{3}{2}^+\rangle$       & 2 &0&0    & 0 & $\frac{3}{2}$ &$\frac{3}{2}^+$ &$^{2}\Psi^{\rho\rho}_{00} \chi^{s}_{S_z}\phi_{\Xi_c^{'}}$ &$\Xi_c\pi$,$\Xi_c\eta$, $\Lambda_cK$,  $\Xi_c'\pi$, $\Sigma_cK$,$\Xi_c^*\pi$, $\Sigma_c^*K$,$\Xi_c(2790,2815)\pi$     \\
\hline
$|\Xi_c^{'}\ ^2 S_{\lambda\lambda}\frac{1}{2}^+\rangle$  & 2 &0&0    & 0 & $\frac{1}{2}$ &$\frac{1}{2}^+$ &$^{2}\Psi^{\lambda\lambda}_{00}\chi^{\lambda}_{S_z}\phi_{\Xi_c^{'}}$ &$\Xi_c\pi$,$\Xi_c\eta$, $\Lambda_cK$,  $\Xi_c'\pi$, $\Sigma_cK$,$\Xi_c^*\pi$, $\Sigma_c^*K$,$\Xi_c(2790,2815)\pi$\\
\hline
$|\Xi_c^{'} \ ^4 S_{\lambda\lambda}\frac{3}{2}^+\rangle$ & 2 &0&0    & 0 & $\frac{3}{2}$ &$\frac{3}{2}^+$ & $^{2}\Psi^{\lambda\lambda}_{00} \chi^{s}_{S_z}\phi_{\Xi_c^{'}}$&$\Xi_c\pi$,$\Xi_c\eta$, $\Lambda_cK$,  $\Xi_c'\pi$, $\Sigma_cK$,$\Xi_c^*\pi$, $\Sigma_c^*K$,$\Xi_c(2790,2815)\pi$\\
\hline
\end{tabular}
\end{table}
\end{center}
\end{widetext}

\section{The chiral quark model }\label{qmc}

In the chiral quark model, the effective low energy quark-meson
pseudoscalar coupling at tree-level is given
by\cite{Li:1995si,Zhong:2007fx,Li:1997gda,qk3,Zhong:2011ti}
\begin{eqnarray}\label{coup}
H_m=\sum_j
\frac{1}{f_m}\bar{\psi}_j\gamma^{j}_{\mu}\gamma^{j}_{5}\psi_j\vec{\tau}\cdot
\partial^{\mu}\vec{\phi}_m.
\end{eqnarray}
where $\psi_j$ represents the $j$-th quark field in a baryon and
$f_m$ is the meson's decay constant. The pseudoscalar-meson octet
$\phi_m$ is expressed as
\begin{eqnarray}
\phi_m=\pmatrix{
 \frac{1}{\sqrt{2}}\pi^0+\frac{1}{\sqrt{6}}\eta & \pi^+ & K^+ \cr
 \pi^- & -\frac{1}{\sqrt{2}}\pi^0+\frac{1}{\sqrt{6}}\eta & K^0 \cr
 K^- & \bar{K}^0 & -\sqrt{\frac{2}{3}}\eta}.
\end{eqnarray}

To match non-relativistic harmonic oscillator spatial wave function
$^N\Psi_{LL_z}$ in the quark model, we adopt the non-relativistic
form of Eq. (\ref{coup}) in the calculations, which is given by
\cite{Li:1995si,Zhong:2007fx,Li:1997gda,qk3,Zhong:2011ti}
\begin{eqnarray}\label{non-relativistic-expans}
H^{nr}_{m}&=&\sum_j\Big\{\frac{\omega_m}{E_f+M_f}\vsig_j\cdot
\textbf{P}_f+ \frac{\omega_m}{E_i+M_i}\vsig_j \cdot
\textbf{P}_i \nonumber\\
&&-\vsig_j \cdot \textbf{q} +\frac{\omega_m}{2\mu_q}\vsig_j\cdot
\textbf{p}'_j\Big\}I_j \varphi_m,
\end{eqnarray}
where $\vsig_j$ and $\mu_q$ correspond to the Pauli spin vector and
the reduced mass of the $j$-th quark in the initial and final
baryons, respectively. For emitting a meson, we have
$\varphi_m=e^{-i\textbf{q}\cdot \textbf{r}_j}$, and for absorbing a
meson we have $\varphi_m=e^{i\textbf{q}\cdot \textbf{r}_j}$. In the
above non-relativistic expansions, $\textbf{p}'_j=\textbf{p}_j-m_j/M
\textbf{P}_{c.m.}$ is the internal coordinate for the $j$-th quark
in the baryon rest frame. $\omega_m$ and $\textbf{q}$ are the energy
and three-vector momentum of the meson, respectively. $\textbf{P}_i$
and $ \textbf{P}_f$ stand for the momenta of the initial final
baryons, respectively. The isospin operator $I_j$ in Eq.
(\ref{non-relativistic-expans}) is expressed as
\begin{eqnarray}
I_j=\cases{ a^{\dagger}_j(u)a_j(s) & for $K^+$, \cr
a^{\dagger}_j(s)a_j(u) & for $K^-$,\cr a^{\dagger}_j(d)a_j(s) & for
$K^0$, \cr a^{\dagger}_j(s)a_j(d) & for $\bar{K^0}$,\cr
a^{\dagger}_j(u)a_j(d) & for $\pi^-$,\cr a^{\dagger}_j(d)a_j(u)  &
for $\pi^+$,\cr
\frac{1}{\sqrt{2}}[a^{\dagger}_j(u)a_j(u)-a^{\dagger}_j(d)a_j(d)] &
for $\pi^0$, \cr
\frac{1}{\sqrt{2}}[a^{\dagger}_j(u)a_j(u)+a^{\dagger}_j(d)a_j(d)]\cos\phi_P
\cr - a^{\dagger}_j(s)a_j(s)\sin\phi_P & for $\eta$,}
\end{eqnarray}
where $a^{\dagger}_j(u,d,s)$ and $a_j(u,d,s)$ are the creation and
annihilation operators for the $u$, $d$ and $s$ quarks, and $\phi_P$
is the mixing angle of $\eta$ meson in the flavor
basis~\cite{Nakamura:2010zzi,Zhong:2011ht}.

For a light pseudoscalar meson emission in a baryon strong decays,
the partial decay amplitudes can be worked out according to the
non-relativistic operator of quark-meson coupling. The details of
how to work out the decay amplitudes can be seen in our previous
work~\cite{Zhong:2007gp}. The quark model permitted two body strong
decay channels of each charm-strange baryon have been listed in
Tabs. \ref{wf1} and \ref{wf2} as well. With the partial decay
amplitudes derived from the chiral quark model, we can calculate the
strong decay width by
\begin{equation}\label{dww}
\Gamma=\left(\frac{\delta}{f_m}\right)^2\frac{(E_f+M_f)|\textbf{q}|}{4\pi
M_i(2J_i+1)} \sum_{J_{iz},J_{fz}}|\mathcal{M}_{J_{iz},J_{fz}}|^2 ,
\end{equation}
where $\mathcal{M}_{J_{iz},J_{fz}}$ is the transition amplitude,
$J_{iz}$ and $J_{fz}$ stand for the third components of the total
angular momenta of the initial and final baryons, respectively.
$\delta$ as a global parameter accounts for the strength of the
quark-meson couplings. It has been determined in our previous study
of the strong decays of the charmed baryons and heavy-light
mesons~\cite{Zhong:2007gp,Zhong:2008kd}. Here, we fix its value the
same as that in Refs.~\cite{Zhong:2008kd,Zhong:2007gp}, i.e.
$\delta=0.557$.

In the calculation, the standard quark model parameters are adopted.
Namely, we set $m_u=m_d=330$ MeV, $m_s=450$ MeV, $m_c=1700$ MeV and
$m_b=5000$ MeV for the constituent quark masses. The harmonic
oscillator parameter $\alpha$ in the wave function $^N\Psi_{LL_z}$
is taken as $\alpha=0.40$ GeV. The decay constants for $\pi$, $K$
and $\eta$ mesons are taken as $f_{\pi}=132$ MeV, $f_K=f_{\eta}=160$
MeV, respectively. The masses of the mesons and baryons used in the
calculations are adopted from the Particle Data
Group~\cite{Nakamura:2010zzi}. With these parameters, the strong
decay properties of the well known heavy-light mesons and charmed
baryons have been described
reasonably~\cite{Zhong:2007gp,Zhong:2010vq,Zhong:2009sk,Zhong:2008kd}.

\section{RESULTS AND DISCUSSIONS}\label{rds}

\subsection{$\Xi_c^*(2645)$}

$\Xi^*_c(2645)$ and $\Xi'_c $ are the two lowest states in the
$\Xi'_c$-type charm-strange baryons. They are assigned to the two
$S$-wave states, $|\Xi'_c \ ^2 S\frac{1}{2}^+\rangle$ and $|\Xi'_c \
^4 S\frac{3}{2}^+\rangle$, respectively. The decay widths of
$\Xi^*_c(2645) \to \Xi_c \pi$ are calculated. The results are listed
in Tab. \ref{tab}, from which we find that our predictions are in
good agreement with the experimental data~\cite{Nakamura:2010zzi},
and compatible with other theoretical
predictions~\cite{Chen:2007xf,Ivanov:1999bk,Aliev:2010ev,
Albertus:2005zy,Tawfiq:1998nk,Cheng:2006dk}.

\begin{widetext}
\begin{center}
\begin{table}[htbp]
\caption{\label{tab} The decay widths (MeV) of the well-established
charm-strange baryons $\Xi_c^*(2645)$, $\Xi_c(2790)$ and
$\Xi_c(2815)$. }
\begin{tabular}{|c|c|c|c|c|c|c|c|c|c|c|c|c}
\hline\hline
                 & Notation                                    & Channel                 & $\Gamma$(ours)  & $\Gamma_{total}$(ours)  & $\Gamma_{total}$\cite{Chen:2007xf}&$\Gamma_{total}$\cite{Ivanov:1999bk}&$\Gamma_{total}$\cite{Aliev:2010ev} &$\Gamma_{total}$\cite{Albertus:2005zy}& $\Gamma_{total}$\cite{Tawfiq:1998nk}& $\Gamma_{total}$\cite{Cheng:2006dk}   & $\Gamma_{total}^{exp}$ \\
 \hline
 $\Xi_c^*(2645)^0$ & $|\Xi_c'\ ^{4}S\frac{3}{2}^+\rangle$       & $\Xi_c^+\pi^-$          &  1.55             &  2.34          & 1.08   & $3.12\pm 0.44$  & $4.2\pm 1.3$ &$3.03\pm0.1$ & 1.88 & $2.8\pm 0.2$         & $<5.5$               \\
                 &                                             & $\Xi_c^0\pi^0$          &  0.79             &                &        &                 &  & &      &                      &                      \\
 $\Xi_c^*(2645)^+$ & $|\Xi_c'\ ^{4}S\frac{3}{2}^+\rangle$       & $\Xi_c^+\pi^0$          &  0.89             &  2.44          & 1.13   & $3.04\pm 0.50$  & $4.2\pm 1.3$ &$3.18\pm0.1$ & 1.81 & $2.7\pm 0.2$         & $<3.1$              \\
                 &                                             & $\Xi_c^0\pi^+$          &  1.55             &                &        &                 &  & &      &                      &                      \\
 $\Xi_c(2790)^+$ & $|\Xi_c\ ^{2}P_\lambda\frac{1}{2}^-\rangle$& $\Xi_c^{'+}\pi^0$       &  0.92             &  2.72          & 9.9    &                 &  & &      & $8.0^{ +4.7}_{-3.3}$ &$<15$                  \\
                 &                                             & $\Xi_c^{'0}\pi^+$       &  1.80             &                &        &                 &  & &      &                      &                      \\
                 &                                             & $\Xi_c^{*}(2645)^0\pi^{+}$&  $4\times10^{-5}$ &                &        &                 &  & &      &                      &                      \\
                 &                                             & $\Xi_c^{*}(2645)^+\pi^{0}$&  $2\times10^{-5}$ &                &        &   &  & &      &                      &                      \\
$\Xi_c(2790)^0$ & $|\Xi_c\ ^{2}P_\lambda\frac{1}{2}^-\rangle$& $\Xi_c^{'+}\pi^-$        &  1.84             &  2.77           & 10.3  &   &  & &      & $8.5^{ +5.0}_{-3.5}$ &$<12$                  \\
                 &                                             & $\Xi_c^{'0}\pi^0$       &  0.93             &                &        &   &  & &      &                      &                      \\
                 &                                             & $\Xi_c^{*}(2645)^+\pi^{-}$&  $4\times10^{-5}$ &                &        &   &  & &      &                      &                      \\
                 &                                             & $\Xi_c^{*}(2645)^0\pi^{0}$&  $2\times10^{-5}$ &                &        &   &  & &      &                      &                      \\
 $\Xi_c(2815)^+$ & $|\Xi_c\ ^{2}P_\lambda\frac{3}{2}^-\rangle$&$\Xi_c^{'0}\pi^+$        & 0.15              & 1.50           &  5.3   &$1.26\pm 0.13$   &  & & 7.67     & $3.4^{ +2.0}_{-1.4}$ & $<3.5$               \\
                 &                                             &$\Xi_c^{'+}\pi^{0}$      & 0.08              &                &        &   &  & &      &                      &                    \\
                 &                                             &$\Xi_c^{*}(2645)^0\pi^{+}$ & 0.83              &                &        &   &  & &      &                      &                    \\
                 &                                             &$\Xi_c^{*}(2645)^+\pi^{0}$ & 0.44              &                &        &   &  & &      &                      &                    \\
  $\Xi_c(2815)^0$& $|\Xi_c\ ^{2}P_\lambda\frac{3}{2}^-\rangle$&$\Xi_c^{'+}\pi^-$        & 0.17              & 1.64           & 5.5    &   &  & &      & $3.6^{ +2.1}_{-1.5}$ & $<6.5$                   \\
                 &                                             &$\Xi_c^{'0}\pi^{0}$      & 0.09              &                &        &   &  & &      &                      &                    \\
                 &                                             &$\Xi_c^{*}(2645)^+\pi^{-}$ & 0.87              &                &        &   &  & &      &                      &                    \\
                 &                                             &$\Xi_c^{*}(2645)^0\pi^{0}$ & 0.51              &                &        &   &  & &      &                      &                    \\
\hline
 \end{tabular}
\end{table}
\end{center}
\end{widetext}

$\Sigma_b$ and $\Sigma_b^*$ ($\Xi_b'$ and $\Xi_b^*$) are
counterparts of $\Xi'_c $ and $\Xi^*_c(2645)$, respectively.
Recently, the improved measurements of the masses and first
measurements of natural widths of the bottom baryon states
$\Sigma_b^{\pm}$ and $\Sigma_b^{*\pm}$ were reported by CDF
Collaboration~\cite{CDF:2011ac}, and a new neutral excited
bottom-strange baryon with a mass $m=5945.0\pm 0.7\pm 0.3\pm 2.7$
MeV was observed by CMS Collaboration~\cite{CMS}. Given the measured
mass and decay mode of the newly observed bottom-strange baryon,
this state most likely corresponds to $\Xi_b^{*0}$ with $J^P=3/2^+$.
As a by-product, we have calculated the strong decays of the bottom
baryons $\Sigma_b^{\pm}$, $\Sigma_b^{*\pm}$ and $\Xi_b^{*0}$. Our
results together with other model predictions and experimental data
have been listed in Tab.~\ref{bott}. From the table, it is seen that
our predictions are in good agreement with the
measurements~\cite{CDF:2011ac,CMS} and the other model predictions~
\cite{Chen:2007xf,Hernandez:2011tx,Hwang:2006df,Guo:2007qu,Limphirat:2010zz}.
It should be pointed out that the strong decay properties of
$\Xi_b^*$ were studied in~\cite{Limphirat:2010zz,Chen:2007xf}, where
a little large mass $m\simeq 5960$ MeV was adopted. With the recent
measured mass of $\Xi_b^{*0}$, the predicted decay widths
in~\cite{Limphirat:2010zz,Chen:2007xf} should be a little smaller
than their previous predictions.

\begin{widetext}
\begin{center}
\begin{table}[htbp]
\caption{\label{bott} The decay widths (MeV) of the ground $S$-wave
bottom baryons $\Sigma_b^{\pm}$, $\Sigma_b^{*\pm}$ and newly
observed $\Xi_b^*(5945)^0$.}
\begin{tabular}{|c|c|c|c|c|c|c|c|c|c|c|c|c}
\hline\hline
                 & Notation                                    & Channel                 & $\Gamma$(ours)    & $\Gamma$\cite{Chen:2007xf}&$\Gamma$\cite{Hernandez:2011tx}&$\Gamma$\cite{Hwang:2006df} &$\Gamma$\cite{Guo:2007qu}& $\Gamma$\cite{Limphirat:2010zz}& $\Gamma_{exp}$ \\
 \hline
$\Sigma_b(5811)^+$      &  $|\Sigma_b\ ^{2}S\frac{1}{2}^+\rangle$      &$\Lambda_b^0\pi^+$  & 5.9    & 3.5    & 6.0  &4.35   & 6.73--13.45 & 4.82,4.94   & $9.7^{+5.0}_{-3.9}$~\cite{CDF:2011ac}                  \\
$\Sigma_b(5816)^-$      &  $|\Sigma_b\ ^{2}S\frac{1}{2}^+   \rangle$   &$\Lambda_b^0\pi^-$  & 6.7    & 4.7    & 7.7  &5.77   & 6.73--13.45 & 6.31,6.49   & $4.9^{+4.2}_{-3.2}$~\cite{CDF:2011ac}                   \\
$\Sigma_b^*(5832)^{+}$  &  $|\Sigma_b\ ^{4}S\frac{3}{2}^{+}\rangle$   &$\Lambda_b^0\pi^+$  & 10.2    & 7.5    & 11.0 &8.50   &10.00--17.74 & 9.68,10.06  & $11.5^{+3.7}_{-3.7}$~\cite{CDF:2011ac}                   \\
$\Sigma_b^*(5835)^{-}$  &  $|\Sigma_b\ ^{4}S\frac{3}{2}^{+}\rangle$   &$\Lambda_b^0\pi^-$  & 11.0    & 9.2    & 13.2 &10.44  &10.00--17.74 & 11.81,12.34 & $7.5^{+3.1}_{-3.2}$~\cite{CDF:2011ac}                  \\
$\Xi_b^*(5945)^0$     &  $|\Xi_b'\ ^{4}S\frac{3}{2}^{+}\rangle$   &$\Xi_b\pi$          & $0.6$    & 0.85   &   ...   &   ...    &   ...          & 1.83,1.85   &  $2.1\pm 1.7$~\cite{CMS}                                        \\
\hline
 \end{tabular}
\end{table}
\end{center}
\end{widetext}

\subsection{$\Xi_c(2790)$ and $\Xi_c(2815)$}

$\Xi_c(2790)$ and $\Xi_c(2815)$ are two relatively well-determined
$P$-wave charm-strange baryons with quantum numbers $J^P=1/2^-$ and
$3/2^-$, respectively. They were observed in the $\Xi_c'\pi$ and
$\Xi_c\pi\pi$ channels, respectively. The Particle Dada Group
suggests they belong to the same SU(4) multiplet as
$\Lambda_c(2593)$ and $\Lambda_c(2625)$,
respectively~\cite{Nakamura:2010zzi}. According to our previous
study, $\Lambda_c(2593)$ and $\Lambda_c(2625)$ can be well explained
with the $|\Lambda_c \ ^2 P_\lambda\frac{1}{2}^-\rangle$ and
$|\Lambda_c \ ^2 P_\lambda\frac{3}{2}^-\rangle$
assignments~\cite{Zhong:2007gp}. Thus, $\Xi_c(2790)$ and
$\Xi_c(2815)$ should correspond to the $\Xi_c$-type excited states
$|\Xi_c \ ^2 P_\lambda\frac{1}{2}^-\rangle$ and $|\Xi_c \ ^2
P_\lambda\frac{3}{2}^-\rangle$, respectively. With these assignments
we have calculated the strong decay properties of $\Xi_c(2790)$ and
$\Xi_c(2815)$, which are listed in Tab~\ref{tab}. Our predicted
widths are in the range of observations~\cite{Nakamura:2010zzi} and
compatible with other theoretical predictions~\cite{Chen:2007xf,
Cheng:2006dk}. On the other hand, $\Xi_c(2790)$ as a dynamically
generated resonance having $J^P=1/2^-$ was also discussed
in~\cite{JimenezTejero:2009vq}.

\begin{figure}[ht]
\centering \epsfxsize=8.0 cm \epsfbox{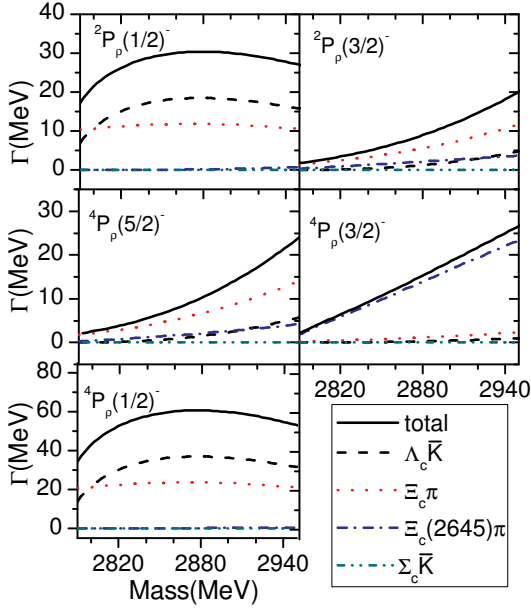} \caption{(Color
online) The strong decay properties of the $P_\rho$-mode excitations
of $\Xi_c$. }\label{fig-p1}
\end{figure}

Finally it should be pointed out that $\Xi_c(2790)$ and
$\Xi_c(2815)$ can not be $P_\rho$-mode excited states $|\Xi_c \ ^2
P_\rho\frac{1}{2}^-\rangle$ and $|\Xi_c \ ^2
P_\rho\frac{3}{2}^-\rangle$, because these excitations have large
partial decay widths into $\Xi_c \pi$ and $\Lambda_c^+ \bar{K}$
channels (see Fig.~\ref{fig-p1}), which disagrees with the
observations. The strong decay properties of the $P_\rho$-mode
excited states have been shown in Fig.~\ref{fig-p1}. We advise
experimentalists to search these missing $P$-wave states in $\Xi_c
\pi$, $\Lambda_c^+ \bar{K}$ and $\Xi_c^*(2645)\pi$ invariant mass
distributions around the energy region $(2.8\sim 2.9)$ GeV.

\subsection{$\Xi_c(2930)$}\label{dsc}

$\Xi_c(2930)$ is not well-established. It was only seen by BaBar in
the $\Lambda_c^+\bar{K}$ invariant mass distribution in an analysis
of $B^-\rightarrow \Lambda_c^+\bar{\Lambda}_c^-K^-$. The mass
analysis of the charm-strange baryon spectrum indicates that
$\Xi_c(2930)$ can be assigned to the first orbital ($1P$) excitation
of $\Xi_c'$ or the first radial ($2S$) excitation of $\Xi_c$ (see
Tab.~\ref{mass2})~\cite{Ebert:2011kk,Ebert:2007nw}.

We have analyze the strong decay properties of all the first
$P$-wave excitations of $\Xi_c'$ and the first radial ($2S$)
excitations of $\Xi_c$, which have been shown in Fig.~\ref{fig-p}
and ~\ref{fig-sa}.

Firstly, we can exclude the first radial ($2S$) excitations of
$\Xi_c$ as assignments to $\Xi_c(2930)$ for the decay channel
$\Lambda_c^+\bar{K}$ of these states is forbidden (see
Fig.~\ref{fig-sa}).

In the first $P$-wave excitations of $\Xi_c'$, we have noted that
the decay modes $\Lambda_c^+\bar{K}$ and $\Xi_c\pi$ for the
$P_\rho$-mode excited states, $^2P_\rho(1/2^-)$ and
$^2P_\rho(3/2^-)$, are forbidden, thus, these states as assignments
to $\Xi_c(2930)$ should be excluded. Furthermore, it is found that
the strong decays of $^2P_\lambda(3/2^-)$ and $^4P_\lambda(5/2^-)$
are governed by the $\Xi_c\pi$ channel, and the $\Xi^*_c(2645)\pi$
decay mode dominates the decay of $^4P_\lambda(3/2^-)$. They might
be hard observed by BaBar for their small $\Lambda_c^+\bar{K}$
branching ratios. Given the decay modes and decay widths, two
$J^P=1/2^-$ states $^4P_\lambda(1/2^-)$ and $^2P_\lambda(1/2^-)$
seem to be the possible assignments to $\Xi_c(2930)$. Considering
$\Xi_c(2930)$ as the $^2P_\lambda(1/2^-)$, from the figure we find
that its decays are dominated by $\Lambda_c^+\bar{K}$ and
$\Xi'_c\pi$, and the other partial decay widths are negligibly
small. Its total width and the partial decay width ratio between
$\Lambda_c^+\bar{K}$ and $\Xi'_c\pi$ are
\begin{eqnarray}
\Gamma=10.6 \mathrm{MeV}, \ \
\frac{\Gamma(\Lambda_c^+\bar{K})}{\Gamma(\Xi'_c\pi)}\approx 1.1.
\end{eqnarray}
And considering $\Xi_c(2930)$ as the $^4P_\lambda(1/2^-)$, we see
that the $\Lambda_c^+\bar{K}$ governs the decays of $\Xi_c(2930)$,
and the other two decay channels $\Xi_c\pi$ and $\Xi'_c\pi$ have
sizeable widths. The calculated total width and partial decay width
ratios are
\begin{eqnarray}
\Gamma=16.7 \mathrm{MeV}, \ \
\frac{\Gamma(\Lambda_c^+\bar{K})}{\Gamma(\Xi_c\pi)}\approx 2.8,\ \
\frac{\Gamma(\Lambda_c^+\bar{K})}{\Gamma(\Xi'_c\pi)}\approx 4.6.
\end{eqnarray}

\begin{widetext}
\begin{center}
\begin{figure}[ht]
\centering \epsfxsize=16.0 cm \epsfbox{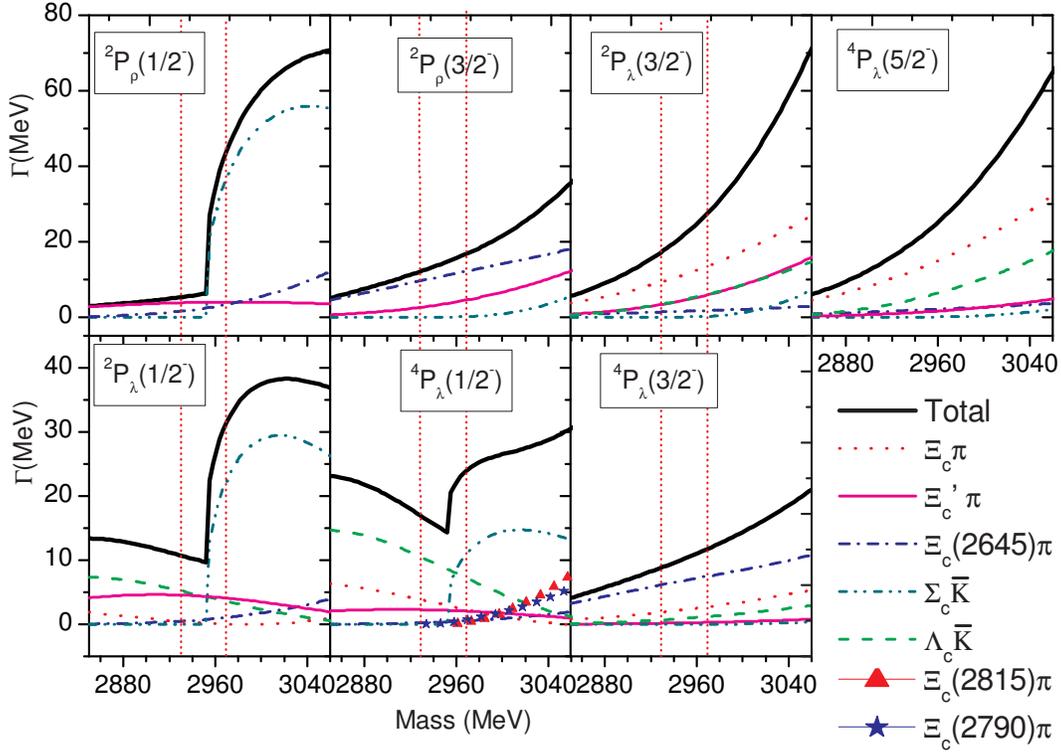} \caption{(Color
online) The strong decay properties of the first orbital ($1P$)
excitations of $\Xi_c'$. }\label{fig-p}
\end{figure}
\end{center}
\end{widetext}

As a whole, $\Xi_c(2930)$ is most likely to be the first orbital
($1P$) excitation of $\Xi_c'$ with $J^P=1/2^-$, favors $|\Xi_c'\
^4P_\lambda 1/2^-\rangle$ or $|\Xi_c'\ ^2P_\lambda 1/2^-\rangle$. To
confirm $\Xi_c(2930)$ and finally classify it, further observations
in the $\Xi'_c\pi$, $\Xi_c\pi$, $\Lambda_c^+\bar{K}$ invariant mass
distributions and measurements of these partial decay ratios are
very crucial in experiments.

\subsection{$\Xi_c(2980)$}\label{2980}

$\Xi_c(2980)$ with a width of $\sim 40$ MeV was first found by Belle
Collaboration in the $\Lambda_c^+\bar{K}\pi$ channel, and then
confirmed by BaBar with large significances in the
intermediate-resonant $\Sigma_c(2455)\bar{K}$ and nonresonant
$\Lambda_c^+\bar{K}\pi$ decay channels. Belle also observed a
resonance structure around $2.97$ GeV with a smaller width of $\sim
18$ MeV in the $\Xi_c^*(2645)\pi$ decay channel in a separate
study~\cite{Lesiak:2008wz}, which is often considered as the same
state of $\Xi_c(2980)$. It should be pointed out that BaBar and
Belle had analyzed the $\Lambda_c^+\bar{K}$ and $\Xi_c\pi$ invariant
mass distributions, respectively, but they did not find any
structures around 2.98 GeV, which indicates that these partial decay
width are too small to be observed or these decay modes are
forbidden. Although $\Xi_c(2980)$ is well-established in
experiments, its quantum numbers are still unknown. Recently, Ebert
\emph{et al.} calculated the mass spectra of heavy baryons in the
heavy-quark-light-diquark picture in the framework of the
QCD-motivated relativistic quark model, they suggested $\Xi_c(2980)$
could be assigned to the first radial ($2S$) excitation of $\Xi_c'$
with $J^P=1/2^+$~\cite{Ebert:2011kk}, which also consists with their
early mass analysis~\cite{Ebert:2007nw}. Cheng \emph{et al.} also
discussed the possible classification of $\Xi_c(2980)$. They
considered that $\Xi_c(2980)$ might be the first radial ($2S$)
excitation of $\Xi_c$ with $J^P=1/2^+$~\cite{Cheng:2006dk}.

\begin{figure}[ht]
\centering \epsfxsize=8 cm \epsfbox{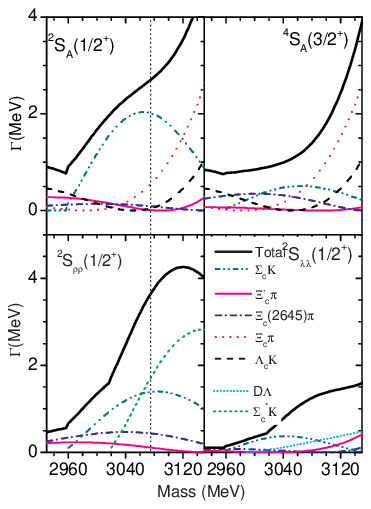} \caption{(Color
online) The strong decay properties of the $L=0$ excitations of
$\Xi_c$. }\label{fig-sa}
\end{figure}

\begin{figure}[ht]
\centering \epsfxsize=8 cm \epsfbox{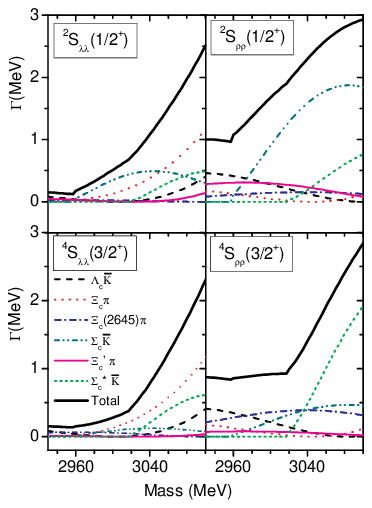} \caption{(Color
online)  The strong decay properties of the first radial ($2S$)
excitations of $\Xi_c'$. }\label{fig-s2}
\end{figure}

We have analyzed the strong decay properties of the first radial
($2S$) excitations of both $\Xi_c$ and $\Xi_c'$, which have been
shown in Figs.~\ref{fig-sa} and~\ref{fig-s2}, respectively. From the
figures, it is seen that the $2S$ excitations of both $\Xi_c$ and
$\Xi'_c$ have narrow decay widths ($< 2$ MeV), which are at least an
order smaller than those of $\Xi_c(2980)$. Furthermore, the decay
modes of these first radial excitations are in disagreement with the
observations of $\Xi_c(2980)$. Thus, the $2S$ excitations of both
$\Xi_c$ and $\Xi'_c$ are excluded as assignments to $\Xi_c(2980)$ in
present work. Our conclusion is in agreement with that of $^3P_0$
calculations~\cite{Chen:2007xf}.

\begin{figure}[ht]
\centering \epsfxsize=8 cm \epsfbox{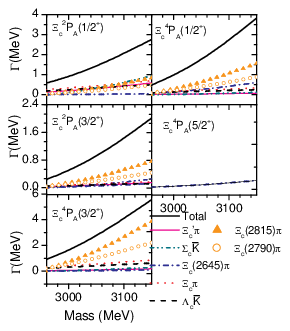} \caption{(Color
online) The strong decay properties of the $P_A$-mode excitations of
$\Xi_c$. }\label{fig-aa}
\end{figure}

$\Xi_c(2980)$ might be one of the $N=2$ shell orbital excitations of
$\Xi_c$ since the masses of these states are close to 2980 MeV.
Their calculated partial decay widths and total widths have been
shown in Figs.~\ref{fig-aa} and~\ref{fig-ds}. It is seen that the
$P_A(1/2^+,3/2^+,5/2^+)$, $D_A(5/2^+,7/2^+)$,
$D_{\rho\rho}(3/2^+,5/2^+)$ and $D_{\lambda\lambda}(3/2^+,5/2^+)$
states have too narrow decay widths to compare with the observations
of $\Xi_c(2980)$. Furthermore, although the decay widths of
$D_A(1/2^+,3/2^+)$ are compatible with the measurement, their decay
modes are dominated by $\Lambda_c^+\bar{K}$ and $\Xi_c\pi$, which
disagrees with the observations as well. As a whole, all the states
shown in Figs.~\ref{fig-aa} and~\ref{fig-ds} are not good
assignments to $\Xi_c(2980)$ either their decay widths are too
narrow to compare with the observations or their decay modes
disagree with the observations.

The $P_\rho$-mode states, $^2P_\rho(1/2^-)$ and $^2P_\rho(3/2^-)$,
in the first $P$-wave excitations of $\Xi_c'$ could be candidates of
$\Xi_c(2980)$ (see Fig.~\ref{fig-p}). We have noted that excitation
of the $\lambda$ variable unlike excitation in $\rho$ involves the
excitation of the ``odd" heavy quark. The $P_\rho$-mode excitation
of charm-strange baryon is $\sim 70$ MeV heavier than the
$P_\lambda$-mode~\cite{Narodetskii:2008pn,Grach:2008ij}. According
to our analysis in \ref{dsc}, $\Xi_c(2930)$ might be assigned to a
$P_\lambda$-mode excitation of $\Xi_c'$. Thus, the expected mass of
the $P_\rho$-mode excitation is $\sim 3.0$ GeV, which is comparable
with that of $\Xi_c(2980)$. As the $^2P_\rho(1/2^-)$ and
$^2P_\rho(3/2^-)$ candidates, respectively, the partial decay widths
and total width of $\Xi_c(2980)$ have been listed in
Tab.~\ref{2980}.

If the resonance structure around $2.97$ GeV in the
$\Xi_c^*(2645)\pi$ decay channel is the same state, $\Xi_c(2980)$,
observed in $\Lambda_c^+\bar{K}\pi$ decay channel, $\Xi_c(2980)$ is
most likely to be the $J^P=1/2^-$ excited state $^2P_\rho(1/2^-)$.
The reasons are as follows. (i) The decay modes of $^2P_\rho(1/2^-)$
are in agreement with the observations. From Tab.~\ref{2980}, we see
that the strong decays of $^2P_\rho(1/2^-)$ are dominated by
$\Sigma_c\bar{K}$, and the partial decay width of $\Xi^*_c(2645)\pi$
is sizeable as well. The $\Lambda_c^+\bar{K}\pi$ final state mainly
comes from a intermediate process in $\Xi_c(2980)\rightarrow
\Sigma_c\bar{K}\rightarrow \Lambda_c^+\bar{K}\pi$. (ii) The total
decay width
\begin{eqnarray}
\Gamma\simeq44 \ \mathrm{MeV},
\end{eqnarray}
is in good agreement with the data. (iii) The decay channels
$\Xi_c\pi$, $\Lambda_c^+\bar{K}$ and $\Sigma^*_c(2520)\bar{K}$ of
$^2P_\rho(1/2^-)$ are forbidden, which can naturally explain why
these decay channels were not observed by Belle and BaBar. It should
be mentioned that the same $J^P$ quantum number (i.e. $J^P$=$1/2^-$)
for $\Xi_c(2980)$ is also suggested in~\cite{JimenezTejero:2009vq},
where the $\Xi_c(2980)$ is considered as a dynamically generated
resonance.

We have noted that the total width of $\Xi_c(2980)$ measured by
Belle and BaBar in the $\Lambda_c^+\bar{K}\pi$ channel is about two
times larger than that measured by Belle in the $\Xi_c^*(2645)\pi$
decay channel in a separate study. Thus, the resonance with a mass
$m\simeq 2970$ MeV [denoted by $\Xi_c(2970)$ in this work] observed
in the $\Xi_c^*(2645)\pi$ decay channel might be a different
resonance from the $\Xi_c(2980)$ observed in the
$\Lambda_c^+\bar{K}\pi$ channel, although they have comparable
masses. According to our analysis, the $\Xi_c(2970)$ observed in the
$\Xi_c^*(2645)\pi$ channel and $\Xi_c(2980)$ observed in the
$\Lambda_c^+\bar{K}\pi$ channel might be assigned to the
$^2P_\rho(1/2^-)$ and $^2P_\rho(3/2^-)$ excitations, respectively.
If the $^2P_\rho(3/2^-)$ is considered as the $\Xi_c(2970)$ observed
in the $\Xi_c^*(2645)\pi$ channel, its total decay width
\begin{eqnarray}
\Gamma\simeq16 \ \mathrm{MeV},
\end{eqnarray}
and dominant decay channel $\Xi_c^*(2645)\pi$ are in good agreement
with the observations (see Tab.~\ref{2980}). Furthermore, it is
interestedly found that when the $^2P_\rho(1/2^-)$ and
$^2P_\rho(3/2^-)$ excitations are considered as the resonances
observed in the $\Lambda_c^+\bar{K}\pi$ and $\Xi_c^*(2645)\pi$,
respectively, we can naturally explain why the width measured in the
$\Xi_c(2645)^*\pi$ channel is about a factor 2 smaller than that
measured in the $\Lambda_c^+\bar{K}\pi$ channel.

In brief, the $\Xi_c(2970)$ observed in the $\Xi_c^*(2645)\pi$ final
state is most likely a different state from the $\Xi_c(2980)$
observed in the $\Lambda_c^+\bar{K}\pi$ final state. The
$\Xi_c(2980)$ and $\Xi_c(2970)$, as two largely overlapping
resonances, favor to be classified as the $|\Xi_c'\ ^2P_\rho
1/2^-\rangle$ and $|\Xi_c'\ ^2P_\rho 3/2^-\rangle$, respectively. Of
course, for the uncertainties of the data we can not exclude the
$\Xi_c(2970)$ and $\Xi_c(2980)$ as the same resonance, which favors
to be assigned to the $|\Xi_c'\ ^2P_\rho 1/2^-\rangle$. To finally
clarify whether the $\Xi_c(2970)$ observed in $\Xi_c^*(2645)\pi$ is
the same resonance observed in the $\Lambda_c^+\bar{K}\pi$ channel
or not, we expect to measure the partial width ratio
$\Gamma[\Xi_c^*(2645)\pi]:\Gamma(\Sigma_c\bar{K})$ further. If there
is only one resonance assigned to $|\Xi_c'\ ^2P_\rho 1/2^-\rangle$,
the ratio $\Gamma[\Xi_c^*(2645)\pi]:\Gamma(\Sigma_c\bar{K})$ might
be $\sim 0.08$. Otherwise, if the $\Xi_c(2980)$ and $\Xi_c(2970)$
corresponds two overlapping resonances $|\Xi_c'\ ^2P_\rho
1/2^-\rangle$ and $|\Xi_c'\ ^2P_\rho 3/2^-\rangle$, respectively,
the ratio might be
$\Gamma[\Xi_c^*(2645)\pi]:\Gamma(\Sigma_c\bar{K})\simeq 0.41$.

\begin{table}[ht]
\caption{The partial decay widths and total width (MeV) for
$\Xi_c(2980)$ as the $^2P_\rho(1/2^-)$ and $^2P_\rho(3/2^-)$
excitations of $\Xi_c'$, respectively. } \label{2980}
\begin{tabular}
{|c|c|c|c|c|c|c|c|c|c|c|c|c|c| }\hline\hline
         & $\Sigma_c\bar{K}$& $\Xi^*_c(2645)\pi$ & $\Xi'_c\pi$ & total   \\
\hline
$^2P_\rho(1/2^-)$ & 37     & 3 & 4     & 44      \\
\hline
$^2P_\rho(3/2^-)$ & 0.1    & 12 & 4   & 16      \\
\hline
\end{tabular}
\end{table}

\subsection{$\Xi_c(3080)$}

$\Xi_c(3080)^+$ and its isospin partner state $\Xi_c(3080)^0$ were
first observed by Belle in the $\Lambda_c^+K^-\pi^+$ and
$\Lambda_c^+K^0\pi^-$ final state, respectively. The existence of
$\Xi_c(3080)^{+,0}$ has been confirmed by BaBar Collaboration.
Furthermore, BaBar's analysis shows that most of the decay of
$\Xi_c(3080)^+$ proceeds through the intermediate resonant modes
$\Sigma_c(2455)^{++}K^-$ and $\Sigma_c(2520)^{++}K^-$ with roughly
equal branching fractions.

Although $\Xi_c(3080)$ has been established in experiments, its
quantum is still unclear. Recently, Ebert \emph{et al.} suggested
$\Xi_c(3080)$ might be classified as the second orbital ($1D$)
excitations of $\Xi_c$ with $J^P=5/2^+$ according to their mass
calculations in the QCD-motivated relativistic quark model. Cheng
\emph{et al.} discussed the possible classification of $\Xi_c(3080)$
as well. They suggested that $\Xi_c(3080)$ might be the second
orbital ($1D$) excitation of $\Xi_c$ with $J^P=3/2^+$ or
$J^P=5/2^+$. More possible assignments to the $\Xi_c(3080)$ were
suggested by Chen \emph{et al.} in their $^3P_0$ strong decay
analysis~\cite{Chen:2007xf}.

\begin{figure}[ht]
\centering \epsfxsize=8 cm \epsfbox{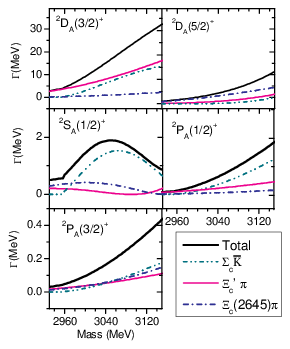} \caption{(Color
online) The strong decay properties of the $l_\lambda=l_\rho=1$
excitations of $\Xi_c'$. }\label{fig-da}
\end{figure}

BaBar's observations provide us two very important constraints on
the assignments to $\Xi_c(3080)$: (i) the strong decay is governed
by both $\Sigma_c(2455)\bar{K}$ and $\Sigma_c(2520)\bar{K}$, (ii)
and the partial width ratio
$\Gamma(\Sigma_c(2455)\bar{K})/\Gamma(\Sigma_c(2520)\bar{K})\simeq
1$. We analyzed the strong decay properties of all the $N=2$ shell
excitations of both $\Xi_c$ and $\Xi'_c$, which were shown in
Figs.~\ref{fig-sa}--\ref{fig-dv}. From the figures we find that only
the $|\Xi_c\ ^2S_{\rho\rho} 1/2^+\rangle$ (i.e., the first radial
($2S$) excitation of $\Xi_c$) satisfies the two constraints of
BaBar's observations at the same time: (i) at $m\simeq 3.08$ GeV the
strong decays of $|\Xi_c\ ^2S_{\rho\rho}1/2^+\rangle$ are dominated
by $\Sigma_c(2455)\bar{K}$ and $\Sigma_c(2520)\bar{K}$, the partial
other two decay modes $\Xi_c^*(2645)\pi$ and $\Xi'_c\pi$ only
contribute a very small partial width to the decay, (ii) and the
predicted partial width ratio between $\Sigma_c(2455)\bar{K}$ and
$\Sigma_c(2520)\bar{K}$ is
\begin{eqnarray}
\frac{\Gamma[\Sigma_c(2455)\bar{K}]}{\Gamma[\Sigma_c(2520)\bar{K}]}\simeq
0.8.
\end{eqnarray}
Furthermore, if the $|\Xi_c\ ^2S_{\rho\rho}1/2^+\rangle$ is
considered as an assignment to $\Xi_c(3080)$, the predicted total
width
\begin{eqnarray}
\Gamma\simeq 4\ \mathrm{MeV}
\end{eqnarray}
is also in good agreement with the measurements.

Finally, it should be point out that as a candidate of
$\Xi_c(3080)$, the mass of $|\Xi_c\ ^2S_{\rho\rho}1/2^+\rangle$
consists with the quark model expectations as well. According to our
analysis in Sec.~\ref{2980}, the $\Xi_c(2980)$ (observed in the
$\Lambda_c^+\bar{K}\pi$ final state) and $\Xi_c(2930)$ could be
assigned to $P_\rho$ and $P_\lambda$-mode excitations of $\Xi_c'$,
respectively. The estimated mass splitting between $P_\rho$ and
$P_\lambda$-mode excitation in the $N=1$ shell is
\begin{eqnarray}
\Delta M\simeq \hbar \omega_\rho-\hbar \omega_\lambda\simeq
(2980-2930)\ \mathrm{MeV}=50 \ \mathrm{MeV}.\label{18}
\end{eqnarray}
With the above relation, we can estimate the mass splitting between
$S_{\rho\rho}$ and $S_{\lambda\lambda}$ excitations in the $N=2$
shell, which is
\begin{eqnarray}
M(S_{\rho\rho})-M(S_{\lambda\lambda})\simeq 2\hbar
\omega_\rho-2\hbar \omega_\lambda\simeq 100 \ \mathrm{MeV}.
\end{eqnarray}
In most of the quark models, the predicted masses for the
$S_{\lambda\lambda}$ excitation of $\Xi_c$ are in the range of
$(2.92\sim 2.99)$ GeV (see Tab.~\ref{mass1}), thus, the mass of
$\Xi_c S_{\rho\rho}$ excitation should be in the range  of
$(3.02\sim 3.09)$ GeV, which is comparable with the mass of
$\Xi_c(3080)$.

As a whole, the mass, decay modes, partial width ratio
$\Gamma(\Sigma_c(2455)\bar{K}):\Gamma(\Sigma_c(2520)\bar{K})$ and
total decay width of $|\Xi_c\ ^2 S_{\rho\rho} 1/2^+\rangle$ strongly
support it is assigned to $\Xi_c(3080)$.

\subsection{$\Xi_c(3055)^+$}\label{3055}

The $\Xi_c(3055)^+$ as a new structure was found by BaBar in the
$\Lambda_c^+\bar{K}\pi$ mass distribution with a statistical
significance of 6.4$\sigma$. It decays through the intermediate
resonant mode $\Sigma_c(2455)^{++}K^-$. BaBar also searched the
inclusive $\Lambda_c^+\bar{K}$ and $\Lambda_c^+\bar{K}\pi\pi$
invariant mass spectra for evidence of $\Xi_c(3055)^+$, but no
significant structure was found. This state has not yet been
confirmed by Belle. According to the calculations of the
charm-strange baryon spectrum in various quark models, $\Xi_c(3055)$
might be assigned to the second orbital ($1D$) excitation of $\Xi_c$
(see Tab.~\ref{mass1}).

We have analyzed the strong decay properties of the second orbital
excitations of $\Xi_c$, which have been shown in Figs.~\ref{fig-aa}
and~\ref{fig-ds}. From Fig.~\ref{fig-aa}, we find that the
$P_A(1/2^+,3/2,5/2^+)$ excitations can be firstly excluded as the
candidates of $\Xi_c(3055)^+$ for neither their decay modes nor
their decay widths consist with the observations. Furthermore, from
Fig.~\ref{fig-ds} it is seen that the $\Lambda_c^+\bar{K}$ is one of
the main decay modes of $^4D_A(1/2^+,3/2,7/2^+)$ and $^2
D_A(3/2^+,5/2^+)$, if the $\Sigma_c(2455)^{++}K^-$ decay mode for
these states is observed in experiments, the $\Lambda_c^+\bar{K}$
decay mode should be observed as well, which disagrees with the
observations of BaBar. Thus, these states as assignments to
$\Xi_c(3055)^+$ should be excluded. The strong decays of $^4
D_A(5/2^+)$, $^2 D_{\lambda\lambda}(5/2^+)$ and $^2
D_{\rho\rho}(5/2^+)$ are dominated by $\Xi_c^*(2645)\pi$ and
$\Sigma_c(2520)\bar{K}$, the partial width of
$\Sigma_c(2455)\bar{K}$ is negligibly small, thus, these states can
not be considered as candidates of $\Xi_c(3055)^+$ as well.

\begin{widetext}
\begin{center}
\begin{figure}[ht]
\centering \epsfxsize=16 cm \epsfbox{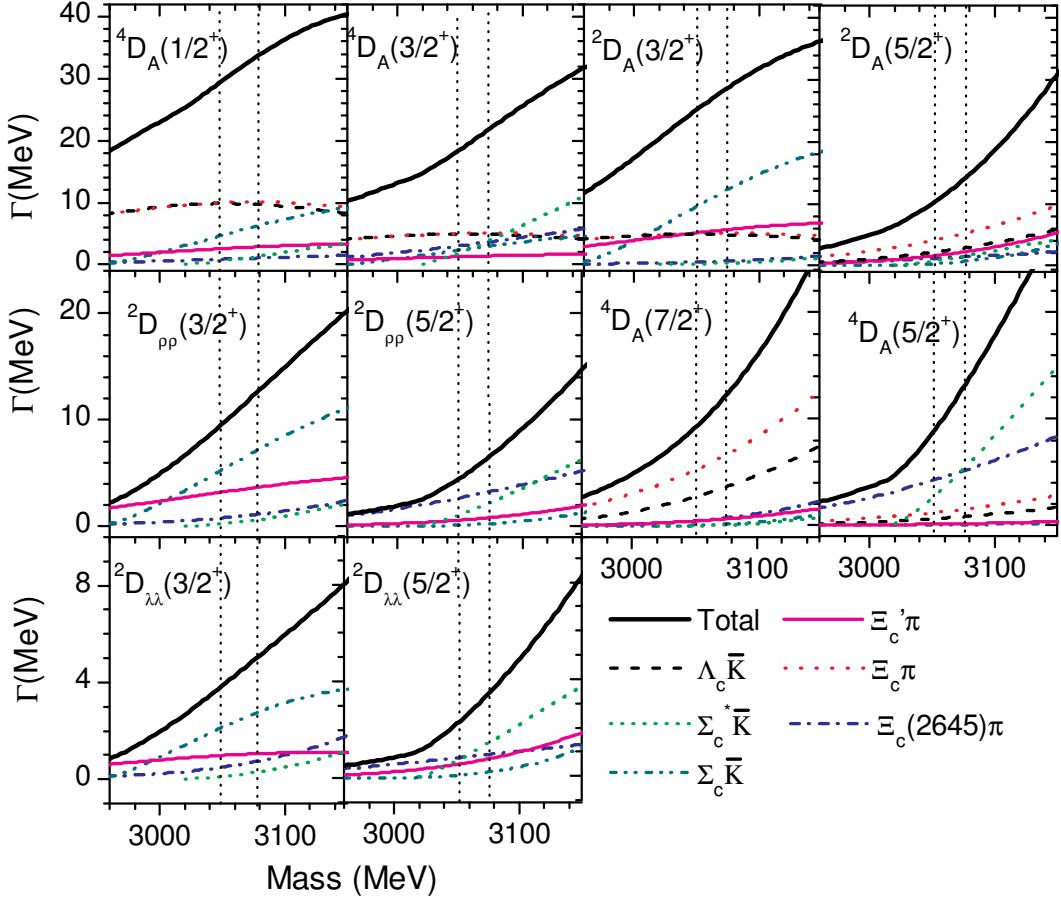} \caption{(Color
online) The strong decay properties of the second orbital ($1D$)
excitations of $\Xi_c$. Some decay channels, such as $\Xi_c\eta$,
$\Xi_c(2790,2815)\pi$ are not shown in the figure for their small
partial decay widths.}\label{fig-ds}
\end{figure}
\end{center}
\end{widetext}

Finally, we find that only two $J^P=3/2^+$ states $|\Xi_c\
^2D_{\lambda\lambda} 3/2^+\rangle$ and $|\Xi_c\ ^2D_{\rho\rho}
3/2^+\rangle$, might be candidates of the $\Xi_c(3055)$. The partial
decay widths and total width of $\Xi_c(3055)$ as the $|\Xi_c\
^2D_{\lambda\lambda} 3/2^+\rangle$ and $|\Xi_c\ ^2D_{\rho\rho}
3/2^+\rangle$ candidates have been listed in Tab.~\ref{3055},
respectively. From the table it is seen that the total widths of
both states are compatible with the observations of $\Xi_c(3055)$
within its uncertainties. The strong decays of both states are
dominated by $\Sigma_c(2455)\bar{K}$ and the partial width of
$\Sigma_c(2520)\bar{K}$ is negligibly small, which can explain why
BaBar only observed the intermediate resonant decay mode
$\Sigma_c(2455)^{++}K^-$ for $\Xi_c(3055)$. The $\Lambda_c^+\bar{K}$
decay mode is forbidden for both  $|\Xi_c\ ^2D_{\lambda\lambda}
3/2^+\rangle$ and $|\Xi_c\ ^2D_{\rho\rho} 3/2^+\rangle$, which
agrees with the observation that no structures were found around
$M(\Lambda_c^+\bar{K})\simeq 3.05$ GeV. As a whole, $\Xi_c(3055)$
could be assigned to the second orbital ($1D$) excitations of
$\Xi_c$ with $J^P=3/2^+$, our conclusion is in agreement with that
of Ebert \emph{et al.} according to their mass analysis. However, it
is difficult to determine which one can be assigned to
$\Xi_c(3055)^+$ in the  $|\Xi_c\ ^2D_{\lambda\lambda} 3/2^+\rangle$
and $|\Xi_c\ ^2D_{\rho\rho} 3/2^+\rangle$ candidates only according
to the strong decay properties. We have noted that $\Xi_c(3080)$ is
most likely to be the $\Xi_c S_{\rho\rho}$ assignment. According to
various quark model predictions, the mass of the second orbital
excitation $\Xi_c D_{\rho\rho}$ should be larger than that of the
first radial excitation $\Xi_c S_{\rho\rho}$, which indicates that
the mass of $\Xi_c D_{\rho\rho}$ might be larger than 3.08 GeV. From
this point of view, the $|\Xi_c\ ^2D_{\rho\rho} 3/2^+\rangle$ as an
assignments to $\Xi_c(3055)^+$ should be excluded. Thus, the
$\Xi_c(3055)$ is most likely to be classified as the $|\Xi_c\
^2D_{\lambda\lambda}3/2^+\rangle$ excitation.

\begin{table}[ht]
\caption{The partial decay widths and total width (MeV) for
$\Xi_c(3055)$ as the $^2D_{\lambda\lambda}(3/2^+)$ and
$^2D_{\rho\rho}(3/2^+)$ excitations of $\Xi_c$, respectively. }
\label{3055}
\begin{tabular}
{|c|c|c|c|c|c|c|c|c|c|c|c|c|c| }\hline\hline
         & $\Sigma_c\bar{K}$& $\Xi^*_c(2645)\pi$ & $\Xi'_c\pi$ & $\Sigma_c^*\bar{K}$&$D\Lambda$ &total   \\
\hline
$^2D_{\lambda\lambda}(3/2^+)$ & 2.3     & 0.5 & 1.0     &0.1 &0.1&4.0      \\
\hline
$^2D_{\rho\rho}(3/2^+)$ & 5.6    & 0.8 & 3.3   & 0.3 &--& 10.0    \\
\hline
\end{tabular}
\end{table}

\subsection{$\Xi_c(3123)^+$}

$\Xi_c(3123)^+$ is another new narrow structure was observed by
BaBar in the $\Lambda_c^+\bar{K}\pi$ final state only with week
statistical significance $3.0\sigma$. It decays through a
intermediate resonant process in
$\Xi_c(3123)^+\rightarrow\Sigma_c(2520)^{++}K^-\rightarrow\Lambda_c^+K^-\pi^+$.
BaBar also searched $\Xi_c(3123)^+$ in the $\Lambda_c^+\bar{K}$ and
$\Lambda_c^+\bar{K}\pi\pi$ final states further, however, they did
not find any evidence in these channels. $\Xi_c(3123)^+$ has not yet
been confirmed by Belle.

From Tab.~\ref{mass1}, it is seen that the predicted masses of the
second orbital ($1D$) excitations of $\Xi_c'$ in various quark
models are $(3.12\sim 3.17)$ GeV. Thus, the $1D$ excitations of
$\Xi_c'$ might be candidates of $\Xi_c(3123)^+$. We have analyzed
the strong decay properties of these excitations, which have been
shown in Fig.~\ref{fig-dv}.

In these $D$-wave states, we can first excluded the
$^2D_{\rho\rho}(3/2^+)$, $^4D_{\rho\rho}(1/2^+)$,
$^2D_{\lambda\lambda}(3/2^+)$, $^2D_{\lambda\lambda}(5/2^+)$,
$^4D_{\lambda\lambda}(1/2^+)$ and $^4D_{\lambda\lambda}(7/2^+)$ as
assignments to $\Xi_c(3123)^+$ for their partial width of
$\Sigma_c(2520)\bar{K}$ is negligibly small compared with that of
$\Sigma_c(2455)\bar{K}$ or $\Lambda_c^+\bar{K}$. Furthermore, we do
not consider the $^2D_{\rho\rho}(5/2^+)$, $^4D_{\rho\rho}(3/2^+)$
and $^4D_{\rho\rho}(7/2^+)$ as good candidates of $\Xi_c(3123)^+$
although the $\Sigma_c(2520)\bar{K}$ is their dominant decay
channel. The reason is that the $\Lambda_c^+\bar{K}$ has a large
partial width which should be observed by BaBar, however, this decay
mode was not observed yet.

Finally, only three excitations $^4D_{\lambda\lambda}(3/2^+)$,
$^4D_{\lambda\lambda}(5/2^+)$ and $^4D_{\rho\rho}(5/2^+)$ might be
candidates of $\Xi_c(3123)^+$. They decay mainly through
$\Sigma_c(2520)\bar{K}$ with a narrow decay width, which is
consistent with the observations of $\Xi_c(3123)^+$. To clearly see
the decay properties of $^4D_{\lambda\lambda}(3/2^+)$,
$^4D_{\lambda\lambda}(5/2^+)$ and $^4D_{\rho\rho}(5/2^+)$, as
candidates of $\Xi_c(3123)^+$ their partial decay widths and total
width have been listed in Tab.~\ref{3123}.

According to our analysis in Sec.~\ref{3055}, $\Xi_c(3055)$ is most
likely to be the $|\Xi_c\ ^2D_{\lambda\lambda}3/2^+\rangle$
excitation. We have noted that the quark model predicted mass of
$\Xi_c'D_{\lambda\lambda}$ is typically $\sim 100$ MeV heavier than
that of $\Xi_cD_{\lambda\lambda}$. Thus, when the $|\Xi_c'\
^4D_{\lambda\lambda}3/2^+\rangle$ or $|\Xi_c'\
^4D_{\lambda\lambda}5/2^+\rangle$ excitation is assigned to the
$\Xi_c(3123)$, the quark model predicted mass $\sim 3.15$ GeV is
compatible with the observation. With the relation of $(\hbar
\omega_\rho-\hbar \omega_\lambda)\simeq 50$ MeV in Eq.(\ref{18}), we
can further estimate the mass splitting between $D_{\rho\rho}$ and
$D_{\lambda\lambda}$ excitations in the $N=2$ shell, which is
\begin{eqnarray}
M(D_{\rho\rho})-M(D_{\lambda\lambda})\simeq 2\hbar
\omega_\rho-2\hbar \omega_\lambda\simeq 100 \
\mathrm{MeV}.\label{rabc}
\end{eqnarray}
Thus, the estimated mass of $|\Xi_c'\ ^4D_{\rho\rho}5/2^+\rangle$ is
$\sim 3.25$ GeV. Obviously, the $|\Xi_c'\
^4D_{\rho\rho}5/2^+\rangle$ could not be considered as a good
assignment to $\Xi_c(3123)$ for its mass is too heavy to compare
with the measurement.

\begin{widetext}
\begin{center}
\begin{figure}[ht]
\centering \epsfxsize=17.0 cm \epsfbox{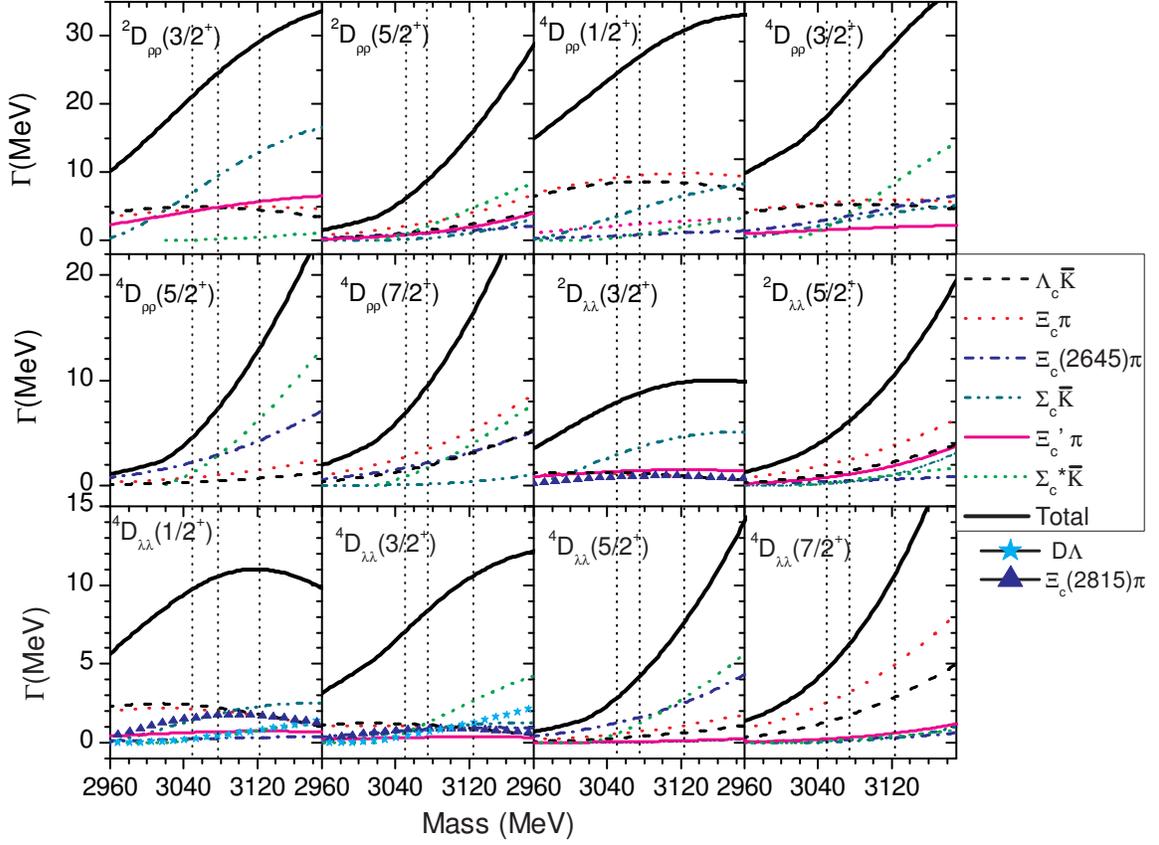} \caption{(Color
online) The strong decay properties of the second orbital ($1D$)
excitations of $\Xi_c'$. Some decay channels, such as $\Xi_c\eta$,
$\Xi_c(2790,2815)\pi$ are not shown in the figure for their too
narrow partial decay widths to compare with the
others'.}\label{fig-dv}
\end{figure}
\end{center}
\end{widetext}

It should be pointed out that in second orbital ($1D$) excitations
of $\Xi_c$, the $D_{\rho\rho}$ excitation $|\Xi_c\ ^2D_{\rho\rho}
5/2^+\rangle$ is also a good assignment to $\Xi_c(3123)$. According
to Eq.~(\ref{rabc}) the mass of the $\rho$ variable excitation
$D_{\rho\rho}$ is $100$ MeV heavier than the $D_{\lambda\lambda}$
excitation. In Sec.~\ref{3055}, we predicted that $\Xi_c(3055)$ is
most likely to be the $|\Xi_c\ ^2D_{\lambda\lambda}3/2^+ \rangle$
excitation, thus, the estimated masses for $|\Xi_c\
^2D_{\rho\rho}5/2^+\rangle$ might be $\sim 3.15$ GeV, which are
close to the mass of $\Xi_c(3123)$. Its partial decay widths and
total width have been listed in Tab.~\ref{3123}. From the table it
is seen that both the decay modes and total width of $|\Xi_c\
^2D_{\rho\rho} 5/2^+\rangle$ are compatible with the observations of
$\Xi_c(3123)$.

As a conclusion, for the scare experimental information, we can not
determine the $J^P$ of $\Xi_c(3123)$. Given the mass, decay mode and
total width observed in experiment, $\Xi_c(3123)$ could be assigned
to the excitation $|\Xi_c'\ ^4D_{\lambda\lambda}3/2^+\rangle$,
$|\Xi_c'\ ^4D_{\lambda\lambda}5/2^+\rangle$ or $|\Xi_c\
^2D_{\rho\rho} 5/2^+\rangle$. Since the $|\Xi_c'\
^4D_{\lambda\lambda} 3/2^+\rangle$, $|\Xi_c'\ ^4D_{\lambda\lambda}
5/2^+\rangle$, and $|\Xi_c\ ^2D_{\rho\rho} 5/2^+\rangle$ have a
comparable mass, the $\Xi_c(3123)$ structure might correspond to
several highly overlapping states around $3.1$ GeV. From
Tab.~\ref{3123}, it is seen that the partial decay width ratios
$\Gamma(\Sigma_c\bar{K}):\Gamma(\Sigma_c^*\bar{K})$,
$\Gamma(\Xi_c^*2645\pi):\Gamma(\Sigma_c^*\bar{K})$ and
$\Gamma(\Xi_c\pi):\Gamma(\Sigma_c^*\bar{K})$ for these possible
assignments to $\Xi_c(3123)$ are very different, thus, the
measurements of these ratios are important to understand the nature
of $\Xi_c(3123)$.

\begin{widetext}
\begin{center}
\begin{table}[ht]
\caption{The partial decay widths and total width (MeV) for
$\Xi_c(3123)$ as the $\Xi_c'\ ^4D_{\lambda\lambda}(3/2^+)$, $\Xi_c'\
^4D_{\lambda\lambda}(5/2^+)$, $\Xi_c'\ ^4D_{\rho\rho}(5/2^+)$ and
$\Xi_c\ ^2D_{\rho\rho}(5/2^+)$ excitations, respectively. }
\label{3123}
\begin{tabular}
{|c|c|c|c|c|c|c|c|c|c|c|c|c|c|c|c }\hline\hline
         & $\Sigma_c\bar{K}$& $\Xi^*_c(2645)\pi$ & $\Xi_c\pi$ & $\Sigma_c^*\bar{K}$& $\Lambda_c\bar{K}$ &$\Xi'_c\pi$&$\Xi_c(2815)\pi$& $\Xi_c(2790)\pi$ &$D\Lambda$ &total
& $\frac{\Gamma(\Sigma_c\bar{K})}{\Gamma(\Sigma_c^*\bar{K})}$& $\frac{\Gamma(\Xi^*_c(2645)\pi)}{\Gamma(\Sigma_c^*\bar{K})}$
& $\frac{\Gamma(\Xi_c\pi)}{\Gamma(\Sigma_c^*\bar{K})}$ \\
\hline
$\Xi_c'\ ^4D_{\lambda\lambda}(3/2^+)$ & 1.2     & 1.0 & 0.9     &2.6& 0.9& 0.4  &0.9 &0.4 &1.2 &10.5 &0.46& 0.38 &0.35   \\
\hline
$\Xi_c'\ ^4D_{\lambda\lambda}(5/2^+)$ & 0.07    & 2.6 & 1.1   & 2.9 & 0.6 & 0.1 & 0.1 &0.05 &0.09 &7.8&0.02&0.90 &0.38  \\
\hline
$\Xi_c'\ ^4D_{\rho\rho}(5/2^+)$ & 0.1    & 4.3 & 1.5   & 6.3 & 0.7 & 0.2 & 0 &0 &0 &13.0 & 0.01&0.68& 0.24 \\
\hline
$\Xi_c\ ^2D_{\rho\rho}(5/2^+)$ & 0.8    & 4.5 & 0   & 4.8 & 0 & 1.5 & 0 &0 &0 &11.6 & 0.17&0.94&0  \\
\hline
\hline
\end{tabular}
\end{table}
\end{center}
\end{widetext}

\section{summary}\label{sumary}

In the chiral quark model framework, the strong decays of
charm-strange baryons are studied. As a by-product we also calculate
the strong decays of the $S$-wave bottom baryons $\Sigma_b^{\pm}$,
$\Sigma_b^{*\pm}$, $\Xi_b'$ and $\Xi_b^*$. We obtain good
descriptions of the strong decay properties of the well-determined
charm-strange baryons $\Xi^*(2645)$, $\Xi(2790)$ and $\Xi(2815)$.
Furthermore, the calculated strong decay widths of $\Sigma_b^{\pm}$,
$\Sigma_b^{*\pm}$, and  $\Xi_b^*$ are in good agreement with the
recent measurements.

$\Xi_c(2930)$, if it could be confirmed in experiments, might be the
first $P$-wave excitations of $\Xi_c'$ with $J^P=1/2^-$. $|\Xi_c'\
^2P_\lambda 1/2^-\rangle$ and $|\Xi_c'\ ^4P_\lambda 1/2^-\rangle$
could be candidates of $\Xi_c(2930)$ according to the present data.
Further observations in the $\Xi'_c\pi$, $\Xi_c\pi$,
$\Lambda_c^+\bar{K}$ invariant mass distributions and measurements
of these partial decay ratios are very crucial to confirm
$\Xi_c(2930)$ and classify it finally.

$\Xi_c(2980)$ might correspond to two different $P_\rho$-mode
excitations of $\Xi_c'$: one resonance is the broader ($\Gamma\simeq
44$ MeV) excitation $|\Xi_c'\ ^2P_\rho 1/2^-\rangle$, which was
observed in the $\Lambda_c^+\bar{K}\pi$ final state by BaBar and
Belle, and the other resonance is the narrower ($\Gamma\simeq 16$
MeV) excitation $|\Xi_c'\ ^2P_\rho 3/2^-\rangle$, which was observed
in the $\Xi_c^*(2645)\pi$ channel by Belle in a separate study. If
the structures were observed in the $\Lambda_c^+\bar{K}\pi$ and
$\Xi_c^*(2645)\pi$ final states correspond to the same state
$\Xi_c(2980)$, which could only be assigned to the $|\Xi_c'\
^2P_\rho 1/2^-\rangle$ excitation. To finally clarify whether the
$\Xi_c(2970)$ observed in $\Xi_c^*(2645)\pi$ is the same state
observed in the $\Lambda_c^+\bar{K}\pi$ channel or not, we expect to
measure the partial width ratio
$\Gamma[\Xi_c^*(2645)\pi]:\Gamma(\Sigma_c\bar{K})$ further.

\begin{widetext}
\begin{center}
\begin{figure}[ht]
\centering \epsfxsize=16.0 cm \epsfbox{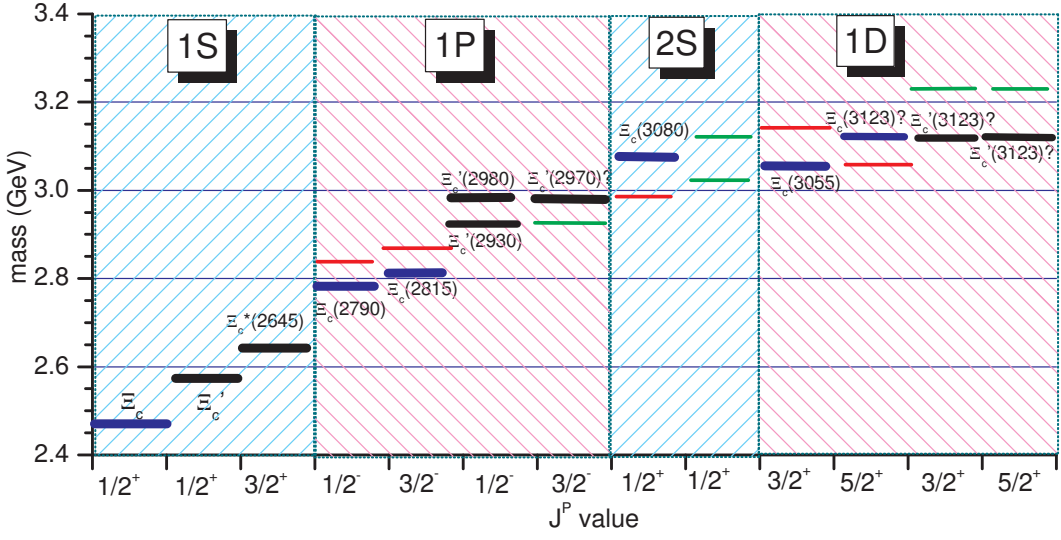} \caption{(Color
online) The charm-strange baryon spectrum up to $N=2$ shell
according to our predictions. In $1P$, $2S$ and $1D$ excitations,
there are two lines for each $J^P$ value, which correspond to the
masses of the excitations of $\rho$ variable (upper line) and
$\lambda$ variable (lower line), respectively. The mass gap between
the $\lambda$ variable excitation and the $\rho$ variable excitation
is assumed to be $50$ MeV for the $1P$ states, and $100$ MeV for the
$2S$ and $1D$ states. The thin lines stand for the states unobserved
in experiments. In the $1P$ ($1D$) excitations, the fist two $J^P$
values are for the excitations of $\Xi_c$, while the last two $J^P$
values are for the excitations of $\Xi_c'$. In $2S$ excitations, the
fist $J^P$ value is for the excitations of $\Xi_c$, while the second
$J^P$ value is for the excitations of $\Xi_c'$. }\label{fig-spectr}
\end{figure}
\end{center}
\end{widetext}

$\Xi_c(3080)$ favors to be identified as the first radial excitation
$|\Xi_c\ ^2S_{\rho\rho}1/2^+\rangle$. The width, decay modes and
ratio
$\Gamma(\Sigma_c(2455)\bar{K})/\Gamma(\Sigma_c(2520)\bar{K})\simeq
0.8$ are in good agreement with the observations. As a assignment to
$\Xi_c(3080)$, the mass of $|\Xi_c\ ^2S_{\rho\rho}1/2^+\rangle$ is
also consistent with the quark model expectations.

Given the mass, decay modes and decay width, $\Xi_c(3055)$ is most
likely to be classified as the second orbital $\Xi_c$ excitation
$|\Xi_c\ ^2D_{\lambda\lambda}3/2^+ \rangle$. To confirm it in
experiments, more observations in the $\Sigma_c\bar{K}$, $\Xi_c'\pi$
and $\Xi_c^*(2645)\pi$ channels are needed.

$\Xi_c(3123)$ is most likely to be the second orbital ($1D$)
excitations of the charm-strange baryon with $J^P=3/2^+$ or $5/2^+$.
It could be assigned to the $\Xi_c'$ excitation $|\Xi_c'\
^4D_{\lambda\lambda} 3/2^+\rangle$ or $|\Xi_c'\ ^4D_{\lambda\lambda}
5/2^+\rangle$. For the scare experiment information about
$\Xi_c(3123)$, we can not exclude it as the assignment to the
$\Xi_c$ excitation $|\Xi_c\ ^2D_{\rho\rho}5/2^+\rangle$. Since the
$|\Xi_c'\ ^4D_{\lambda\lambda} 3/2^+\rangle$, $|\Xi_c'\
^4D_{\lambda\lambda} 5/2^+\rangle$ and $|\Xi_c\ ^2D_{\rho\rho}
5/2^+\rangle$ have a comparable mass, the $\Xi_c(3123)$ structure
might correspond to several largely overlapping resonances. To good
understand $\Xi_c(3123)$ structure, further observations in the
$\Xi^*_c(2645)\pi$, $\Sigma_c^*\bar{K}$ and $\Xi_c\pi$ channels are
expected.

Finally, according to our predictions we establish a spectroscopy
for the observed charm-strange baryons, which is shown in
Fig.~\ref{fig-spectr}. We also estimate the masses of the
charm-strange baryons with different variable ($\lambda$ or $\rho$)
excitation from these newly observed states in experiments, which
are given in Fig.~\ref{fig-spectr}. These missing states might be
found in future experiments. To provide helpful information for
search for the missing charm-strange baryons, in
Figs.~\ref{fig-p1}--\ref{fig-dv} our predictions of their strong
decay properties have been shown as well.



\section*{  Acknowledgements }

This work is supported, in part, by the National Natural Science
Foundation of China (11075051), Program for Changjiang Scholars and
Innovative Research Team in University (IRT0964), the Program
Excellent Talent Hunan Normal University, and the Hunan Provincial
Natural Science Foundation (11JJ7001).


\end{document}